\documentclass[lettersize,journal]{IEEEtran}
\usepackage{amsmath,amsfonts}
\usepackage{algorithmic}
\usepackage{algorithm}
\usepackage{array}
\usepackage[caption=false,font=footnotesize]{subfig}
\usepackage{textcomp}
\usepackage{stfloats}
\usepackage{url}
\usepackage{verbatim}
\usepackage{graphicx}
\usepackage{cite}
\usepackage[numbers,sort&compress]{natbib}

\usepackage[dvipsnames]{xcolor}

\usepackage{url}
\usepackage{booktabs}       
\usepackage{amsfonts}       
\usepackage{nicefrac}       
\usepackage{microtype}      
\usepackage{xcolor}         
\usepackage{hyperref}
\usepackage{orcidlink}
\usepackage{booktabs} 
\usepackage{graphicx}
\usepackage{mathtools}
\usepackage{amsthm}
\usepackage{microtype}
\usepackage{algorithm, algorithmic}
\usepackage{caption,  multirow}
\usepackage{color}
\usepackage{comment}
\usepackage{times}
\usepackage{makecell}
\usepackage{epsfig}
\usepackage{graphicx}
\usepackage{amsmath}
\usepackage{amssymb}
\usepackage{colortbl}
\usepackage{bbm}
\usepackage{pifont}
\usepackage{threeparttable}

\usepackage[export]{adjustbox}
\usepackage[capitalize]{cleveref}
\usepackage[switch,mathlines]{lineno}
\usepackage{xcolor}

\renewcommand\thelinenumber{\makebox[0pt][r]{\ifnum\value{linenumber}<100 0\fi\ifnum\value{linenumber}<10 0\fi\arabic{linenumber}}}

\makeatletter
\g@addto@macro\UrlBreaks{\do\-\do\_\do\/\do\.}
\makeatother

\usepackage{amsmath,amsfonts,bm}

\def\eqref#1{equation~\ref{#1}}

\def\1{\bm{1}}

\DeclareMathAlphabet{\mathsfit}{\encodingdefault}{\sfdefault}{m}{sl}
\SetMathAlphabet{\mathsfit}{bold}{\encodingdefault}{\sfdefault}{bx}{n}

\begin{document}

\title{Spectrum Aware Illumination Estimation Using Multispectral Image}

\author{Hyejin Oh\,\orcidlink{0009-0006-4615-3362}, Woo-Shik Kim\,\orcidlink{0009-0004-8804-9702}, Sangyoon Lee\,\orcidlink{0000-0001-7019-2114}, YungKyung Park\,\orcidlink{0000-0002-8152-0563}, Je-Won Kang\,\orcidlink{0000-0002-1637-9479}
\thanks{Manuscript received XXXX XX, 2025; revised XXXX XX, 2025.}
\thanks{The work was supported by Samsung Electronics and the National Research Foundation of Korea(NRF) grant
funded by the Korea government(MSIT) (RS-2025-23524046).}
\thanks{H. Oh is with the Department of Electronic and Electrical Engineering, Ewha W. University, Seoul, Korea, 03760. (e-mail: hyejin5@ewha.ac.kr).}
\thanks{W.-S. Kim is with Telechips, Gyeonggi-do, Korea, 13453. This work was done while he was with the Samsung Advanced Institute of Technology. (e-mail: jacob.kim@telechips.com).}
\thanks{S. Lee is with the Samsung Advanced Institute of Technology, Gyeonggi-do, Korea, 16678. (e-mail: sangy00n.lee@samsung.com).}
\thanks{Y. Park is with the Department of Design, Ewha W. University, Seoul, Korea. (e-mail: yungkyung.park@ewha.ac.kr).}
\thanks{J.-W. Kang is with the Department of Electronic and Electrical Engineering, Ewha W. University, Seoul, Korea. (e-mail: jewonk@ewha.ac.kr).}
\thanks{Corresponding author: J.-W. Kang.}
\thanks{Copyright © 2026 IEEE. Personal use of this material is permitted. However, permission to use this material for any other purposes must be obtained from the IEEE by sending an email to pubs-permissions@ieee.org. Digital Object Identifier (DOI): 10.1109/TCSVT.2026.3701975}
}

\markboth{Journal of \LaTeX\ Class Files,~Vol.~14, No.~8, August~2026}%
{Shell \MakeLowercase{\textit{et al.}}: A Sample Article Using IEEEtran.cls for IEEE Journals}


\maketitle
\begin{abstract}

Multispectral (MS) imaging extends beyond conventional RGB imaging by capturing more spectral bands, thereby improving illuminant spectrum estimation (ISE). However, existing methods often fail to fully exploit spectral information, resulting in suboptimal performance under diverse lighting conditions and across different sensor domains. Hence, we propose a deep learning framework with a spatio-spectral feature extraction block, which incorporates spectral attention mechanisms to enhance spectral correlation and preserve illuminant-relevant spatial features. Through the inclusion of an illuminant prior (IP), our approach prioritizes specific channels that provide more meaningful information in an MS image. We also propose a  spectral-domain transform across different MS sensor spaces. The results demonstrate that illuminant spectra learned in high-dimensional sensor spaces can be effectively transformed to various lower-dimensional camera sensor spaces without any additional training. To facilitate evaluation, we introduce a real-world MS dataset containing high-dimensional ground-truth illumination spectra captured under diverse lighting conditions. Through extensive experiments, we demonstrate that our method achieves superior accuracy compared to existing models, thus providing a practical solution for real-world ISE. The code and dataset are available at \textit{\url{https://github.com/hyejin5/Spectrum-Aware-Illumination-Estimation-Using-Multispectral-Image}}.

\end{abstract}    
\section{Introduction}
\label{sec:intro}

Though RGB sensors have been the mainstreams for the imaging devices with high resolution and sensitivity, they can capture only a limited range of spectral bands across the electromagnetic spectrum and struggle to handle complex lighting conditions~\cite{lou2015color, bianco2015color}. To address these problems, multispectral (MS) image sensors, which can acquire a broader range of spectral bands than RGB sensors, have been used to obtain detailed spectral information about a scene or objects. Due to these advantages, MS imaging has enabled applications across various fields, including image classification, super-resolution, remote sensing, autonomous driving, and industrial inspection~\cite{li2023spatial, qing2021improved, wang2019spatial,basterretxea2021hsi,aryal2021mobile}. 
Despite the fact that such MS sensors can provide abundant spectral information, the widespread adoption of hyperspectral imaging has been limited by its high production costs. Recently, however, MS sensors have become more cost-effective, thereby facilitating the acquisition of high-resolution images \cite{glatt2024beyond}.


Illuminant spectrum estimation (ISE) aims to recover the spectral power distribution (SPD) of the scene illumination from captured MS images. Because the SPD is closely related to the chromatic characteristics of illumination, it directly affects MS pixel values and downstream processes that rely on stable spectral measurements. This is particularly relevant in real-world pipelines. For example, in camera image signal processor (ISP), reliable estimation of illuminant chromaticity supports auto white balance (AWB) and consistent color rendering \cite{li2025multi, ccgoeswrong, gijsenij2011computational,mosny2006multispectral}. In industrial inspection, color correction improves inspection accuracy \cite{kwon2014critical, yuan2025multispectral, lam2004robust}. In spectral-imaging applications such as in medical and microscopy imaging, illumination variation can bias quantitative analysis \cite{peng2017basic, ljosa2009introduction}.

\begin{figure}
    \centering
    \includegraphics[width=\columnwidth]{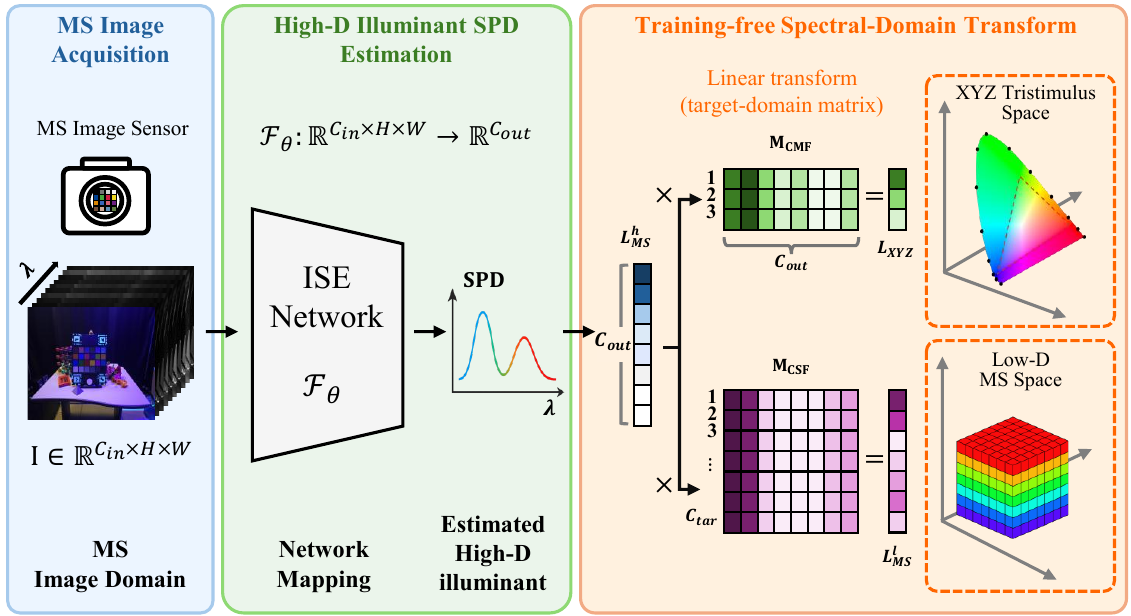}
    \caption{Overall pipeline of the proposed framework. The ISE network estimates a high-dimensional illuminant SPD from an MS image, and the estimated SPD is linearly projected to multiple target domains using target-domain matrices, such as CMF for the XYZ tristimulus space and CSF for the low-dimensional MS sensor space.}
    \label{fig:overall_pipeline}
\end{figure}

Recent studies~\cite{glatt2024beyond,erba2024rgb,erba2024improving} have demonstrated that the increasing number of spectral bands enables more accurate reconstruction of an illuminant SPD function, thereby improving illumination estimation. Various illumination estimation methods based on spectral images have been developed~\cite{zheng2015illumination, khan2018multispectral, khan2018towards, pixelwise, deepunrollingmultispectral, glatt2024beyond}. However, achieving effective ISE remains a challenge, since an illumination estimation problem involves solving an under-constrained problem that requires the simultaneous recovery of both the illumination spectrum and surface reflectance~\cite{zheng2015illumination,stat-spatiospectral,hurlbert2019challenges}. Studies have attempted to impose several constraints on MS images and have used low-rank matrix factorizations to separate scene illumination from surface reflectance~\cite{an2015illumination, zheng2015illumination, chen2017illumination}. However, when the number of channels is less than that of basis functions, the recovery of scene illuminants becomes unstable. The intrinsic interdependence between illumination and surface reflectance spectra hinders accurate illumination estimation of spectral images, especially in complex spectral scenes~\cite{kim2024attentive,zhao2023multi,kinli2023deterministic}.

Apart from the challenge in intrinsic image decomposition, MS images possess inherent complexities in terms of channel variations within the spectral domain. The dimensions of a ground-truth (GT) illuminant spectrum can vary with the channel-sampling intervals and sensor characteristics. Various deep learning techniques have been developed for RGB~\cite{lou2015color, bianco2015color, bianco2017single, FC4, shi2016deep,C42020, clcc} and spectral images~\cite{deepunrollingmultispectral,glatt2024beyond, pixelwise, gu2014segmentation}. However, the aforementioned dimensional variation creates a unique challenge in illuminant estimation, particularly when applying a pre-trained network using one sensor to a dataset acquired using another sensor.
Therefore, for practical applications, it is necessary to develop a spectral-domain transform for illuminant estimation that can accommodate these spectral variations without retraining of a network for each specific case.

In this paper, we propose a comprehensive framework for ISE using MS images, as illustrated in~\cref{fig:overall_pipeline}. 
We present a novel deep learning framework that efficiently utilizes rich spectral information in a spectral image to achieve accurate ISE. Existing techniques operating in the spatial domain treat all channels equally, which result in a loss of spectral correlation among channels, thus leading to a limited capacity to capture spectral information~\cite{li2023spatial,qing2021improved}. Existing models built on a standard transformer model \cite{ViT} often fail to prioritize specific channels that provide more meaningful information. To overcome these drawbacks and leverage the information of the images more effectively, on top of the baseline, the proposed network uses a spatio-spectral feature extraction  block incorporating the spectral attention block with illuminant prior (SABIP) and the multi-head spectral self-attention block (MSSAB). These attention blocks are used to exploit inter-channel relations in the spectral images and preserve spectral information with respect to the illuminant relevant spatial features.  

Furthermore, we propose a spectral-domain transformation method that considers the relationships between both the input image channels ($C_{in}$) and the output illuminant channels ($C_{out}$). This consideration enables the development of a flexible framework that can effectively handle various spectral configurations while maintaining estimation accuracy. Thus, illuminant information learned in higher-dimensional spectral domains can be effectively utilized in lower-dimensional domains through simple transform matrices. The proposed technique produces accurate estimations without requiring additional training procedures. Thus, by generalizing a single model to different MS cameras and conventional trichromatic spaces, encompassing various sensor spaces, practical benefits and versatile solutions can be obtained for real-world illumination estimation problems.

To facilitate this study, we introduce a real-world MS image dataset  MILD (Multispectral Image dataset with Lighting
Diversity), which includes scenes under natural outdoor sunlight, a fluorescent lamp, or a daylight series synthesized using a multi-channel LED lighting booth. Our dataset also includes scenes under extreme lighting conditions, including high chromaticity color illuminants, which has not been examined in other studies. We placed more than two references inside each scene---a color chart, 3D gray, and white. The acquisition was performed using the MS image sensor, fabricated on top of the CMOS image sensor (CIS) for a smartphone, which provides 16-channel information. The GT spectrum for all illuminants is measured using a spectroradiometer resampled to 36 channels in 10 nm intervals across 380-730 nm. The GT data from the spectroradiometer enables the model to learn higher-dimensional illumination spectra.

We conduct extensive experiments on multiple benchmarks, including our dataset containing challenging monochromatic illumination cases that are difficult to handle with RGB images, and provide ablation analyses to validate the effectiveness of each proposed component. The proposed framework achieves state-of-the-art performance across these settings.

The key contributions of this study are as follows:

\begin{itemize}

\item \noindent\textbf{Method: spatio–spectral modeling for MS ISE. } We propose a novel spectral attention-based deep learning framework for ISE from MS images. The proposed design explicitly models spectral inter-channel dependencies via spectral attention, while preserving spectral details with dedicated modules. Our methods enable effective extraction of illuminant cues from complex MS measurements.
\item \noindent\textbf{Generalization: training-free spectral-domain transfer. }We introduce a training-free sensor-space transformation that maps the estimated high-dimensional illuminant spectrum to lower-dimensional target spaces (e.g., MS sensor spaces or XYZ tristimulus space) without additional training. This provides a practical approach to deploy a single ISE model across different sensor configurations.
 
\item \noindent\textbf{Dataset: MS data with spectroradiometer-measured GT and extreme lighting.} We present MILD, a MS image dataset with spectroradiometer-measured ground-truth illuminant spectra, covering indoor illuminants and natural daylight as well as extreme mono-wavelength lighting conditions. This design enables robustness evaluation under extreme illuminant distributions and supports assessing applicability across diverse illumination scenarios.

\end{itemize}

\section{Related Work}
\label{sec:Related Work}

\subsection{Illumination Estimation}
\subsubsection{Statistics-based Illumination Estimation}
Early studies in this domain focused on estimating illumination from RGB images. Statistics-based and deep learning-based methods have been developed to improve the accuracy of illumination estimation. Statistics-based methods~\cite{grey-world,Max-RGB,shadeofGrey,grey-edge,brainard1997bayesian,stat-spatiospectral,land1985retinex} involve modeling techniques that utilized prior information derived from human visual characteristics, physical properties, and statistical observations. The Max-RGB algorithm ~\cite{Max-RGB} assumes the existence of at least one white patch that reflects the perfect illuminant in an image to estimate illumination from the pixel with the maximum spectral response. The Gray-World (GW) algorithm ~\cite{grey-world} assumes that the average reflectance of surfaces in a scene which contains diverse objects is achromatic, similar to gray. The Gray-Edge (GE) algorithm ~\cite{grey-edge} extends these assumptions to image derivatives, assuming that the average edge difference in a scene is achromatic; by comparison, the Shades-of-Gray algorithm ~\cite{shadeofGrey} follows a more generalized assumption, implementing the Minkowski norm as a balance between the GW and Max-RGB methods.
While these statistical methods are computationally efficient, they often underperform in particular scenes that deviate from their underlying assumptions. 

In recent studies, well-established statistical models and relevant assumptions have been extended from RGB to spectral domains \cite{thomas2015illuminant, khan2017illuminant, an2015illumination, chen2017illumination}. 
In ~\cite{khan2017illuminant}, traditional statistics-based illumination estimation algorithms, such as Max-RGB, GW, and GE are extended from RGB to higher dimensional spectral data. In \cite{an2015illumination, zheng2015illumination, chen2017illumination}, scene illumination is separated from surface reflectance via low-rank matrix factorizations. This approach enables a more intricate decomposition of the scene, isolating illumination effects from object colors based on their spectral attributes.

\subsubsection{Deep Learning-based Illumination Estimation}
Deep learning-based methods have been employed to achieve accurate illumination estimation, particularly in complex scenes, such as those with highly specular materials, unusual color distributions, extreme lighting conditions, or limited color diversity where traditional statistical approaches often fail. 
Such methods~\cite{lou2015color, bianco2015color, bianco2017single, FC4, shi2016deep,C42020, clcc} have significantly improved accuracy in complex scenarios by leveraging the spatial characteristics of RGB images. In~\cite{zhang2024color}, complex scene contents are decoupled from illumination using semantic priors in the RGB color domain.

Deep learning-based methods have also been extended to the MS domain for accurate ISE ~\cite{deepunrollingmultispectral,glatt2024beyond, pixelwise, gu2014segmentation}. In ~\cite{deepunrollingmultispectral}, ISE was formulated as a constrained matrix factorization problem, which was optimized using a deep unrolling network to obtain a solution. In ~\cite{pixelwise}, a convolutional neural network (CNN) was employed to estimate pixel-wise scene illuminants by processing image patches at multiple scales; however, it struggled to fully exploit the channel and spectral information in MS images. The unsupervised segmentation method in ~\cite{gu2014segmentation} can model illumination as a piecewise-constant spectrum but fails to handle diverse lighting conditions. In ~\cite{glatt2024beyond}, a pyramid-based convolutional block was designed to process spectral images via triplet decomposition. However, research on ISE using MS images remains limited. The method used in ~\cite{deepunrollingmultispectral} rely on spectral reflectance images, each of which must be calibrated using a reference or measurement device~\cite{li2018efficient}. The limited availability of GT reflectance data under diverse lighting conditions compromises the effectiveness of these aforementioned methods in practical applications. The networks proposed in ~\cite{pixelwise, glatt2024beyond, gu2014segmentation} cannot fully utilize the rich spectral characteristics of MS images. These models often fail to leverage image features relevant to illuminant properties, thereby struggling to accommodate diverse illumination spectra.

Illumination estimation methods have evolved along different trajectories. While statistics-based methods leverage low-level features, including image statistics and spectral information, deep learning-based approaches focus on exploiting data-driven spatial information in images. 
Moreover, recent studies have increasingly employed illumination estimation as an auxiliary module in various tasks, ranging from low-level vision such as low-light enhancement ~\cite{zhu2024ghost}, texture enhancement ~\cite{wu2020illuminance}, and image denoising ~\cite{han2019canonical} to high-level vision tasks such as person re-identification ~\cite{zhang2022illumination}. In \cite{zhu2024ghost}, low-light enhancement is casted as a re-imaging process for objects in dark scenes, in which the illumination estimation is used for a light modulation network. In \cite{han2019canonical}, artifact reduction is addressed under spatially varying illumination conditions. They decompose mixed illumination into mixture weights over a predefined canonical illuminant basis. \cite{wu2020illuminance} compensate illuminance drift as a scalar harmonic field via Hodge decomposition. While these approaches advance illumination modeling in the RGB color space, they do not target wavelength-resolved illuminant SPD estimation from high-dimensional MS measurements.
Deep learning-based ISE frameworks can effectively leverage the rich spectral information contained within MS images while maintaining the advantages of data-driven feature learning.

\subsection{Spectral Information Processing in MS images}

Deep learning-based methods exploit various attention mechanisms to leverage spectral information effectively. The method proposed in \cite{roy2020lightweight} uses the squeeze and excitation (SE) blocks~\cite{SENet}, which recalibrate feature maps by explicitly modeling interdependencies between channels. Likewise, the technique from ~\cite{wang2019spatial} utilizes SE blocks to adaptively learn the weights for different spectral bands and for different neighboring pixels.
The self-attention mechanism, introduced in transformers ~\cite{transformer}, has also been actively used to extract both spectral and spatial features. It effectively captures global long-range dependencies without relying on spatial and temporal locality. Accordingly, spectral self-attention mechanisms have been utilized for various tasks involving spectral images. For image denoising, \cite{li2023spatial} used a spatio-spectral transformer to fully exploit both the non-local spatial similarity and global spectral correlation in MS images. Additionally, \cite{qing2021improved} used the spectral self-attention mechanism to extract spectral–spatial features for image classification. 
Furthermore, to handle cross-modality fusion in MS images, a reverse spectral cross-attention mechanism was employed in ~\cite{chen2024cyclic} to model the differences and similarities between spectral priors and fused features, and a channel-wise fusion attention strategy was adopted in ~\cite{yang2024multidimensional} to capture modality-specific spectral dependencies. 
Recent approaches have focused on combining the strengths of CNNs and transformers into hybrid architectures ~\cite{ma2022hyperspectral, yan2023hyformer}. Such hybrid models use CNNs to extract local spatial features while employing transformer modules to capture global spectral relationships. This combination enables for more comprehensive feature learning, whereby local spatial details are preserved while long-range spectral dependencies are effectively modeled.

While the aforementioned spectral-information-processing architectures have shown promising results in various vision tasks, their application to illumination estimation remains largely unexplored. Accurate illumination estimation necessitates the designing of task-specific architectures that can leverage both spatial and spectral information effectively by capturing illumination-relevant features from the rich spatio-spectral information available in MS images.

\subsection{Spectral Image Datasets for Illumination Estimation}
\subsubsection{Dataset for ISE}
For illumination estimation or tasks related to color processing, numerous previous studies have identified appropriate characteristics of datasets, which have been developed in various formats. Many studies have highlighted the strong dependency of ISE results on the datasets. Bianco et al. developed datasets for several scenarios, including indoor and outdoor lighting ~\cite{bianco2008improving} and single and multiple illuminants ~\cite{bianco2017single}, as well as datasets containing facial imagery ~\cite{bianco2012color}. Busselli et al. investigated ~\cite{buzzelli2023analysis} how image contents and luminance levels affect the performance of common ISE models, observing that images with diverse objects and well-known subjects are more effective for algorithm evaluation. 
In addition, various reference materials have been used to increase the accuracy of illuminant estimation in the image. Gehler et al. ~\cite{gehler2008bayesian} and Cheng et al. \cite{cheng2014illuminant} used the color charts as a reference, while Ciurea \cite{ciurea2003large} and Ershov ~\cite{ershov2020cube++} used 3-D gray references.

\subsubsection{Dataset for Spectral Imaging}
Numerous spectral image datasets have been developed for computer vision applications ~\cite{arad2022ntire, foster2016time, foster2006frequency,nascimento2016spatial, deepunrollingmultispectral, nascimento2002statistics, nguyen2014training, yasuma2010generalized}. The rich spectral information in these datasets offers significant advantages for research, as adaptive channel selection through post-processing can be performed to meet various experimental targets.

However, existing spectral datasets are inherently constrained by their long acquisition times ~\cite{glatt2024beyond}, with the majority being confined to static scenes with limited dynamic content. Furthermore, while spectral reflectance datasets without any illuminant involved would allow illumination manipulation through post-processing, they struggle to accurately represent complex real-world lighting conditions, including inter-reflections, specularity, and variations in spatial illumination.

These observations highlight the need for comprehensive spectral datasets that contain diverse real-world scenes and complex lighting conditions while providing accurate GT illumination information---addressing the current limitations in dataset availability for ISE research.

\section{Proposed Method}
\label{sec:Proposed Method}

\begin{figure*}[ht!]
\centerline{\includegraphics[width=0.95\textwidth]{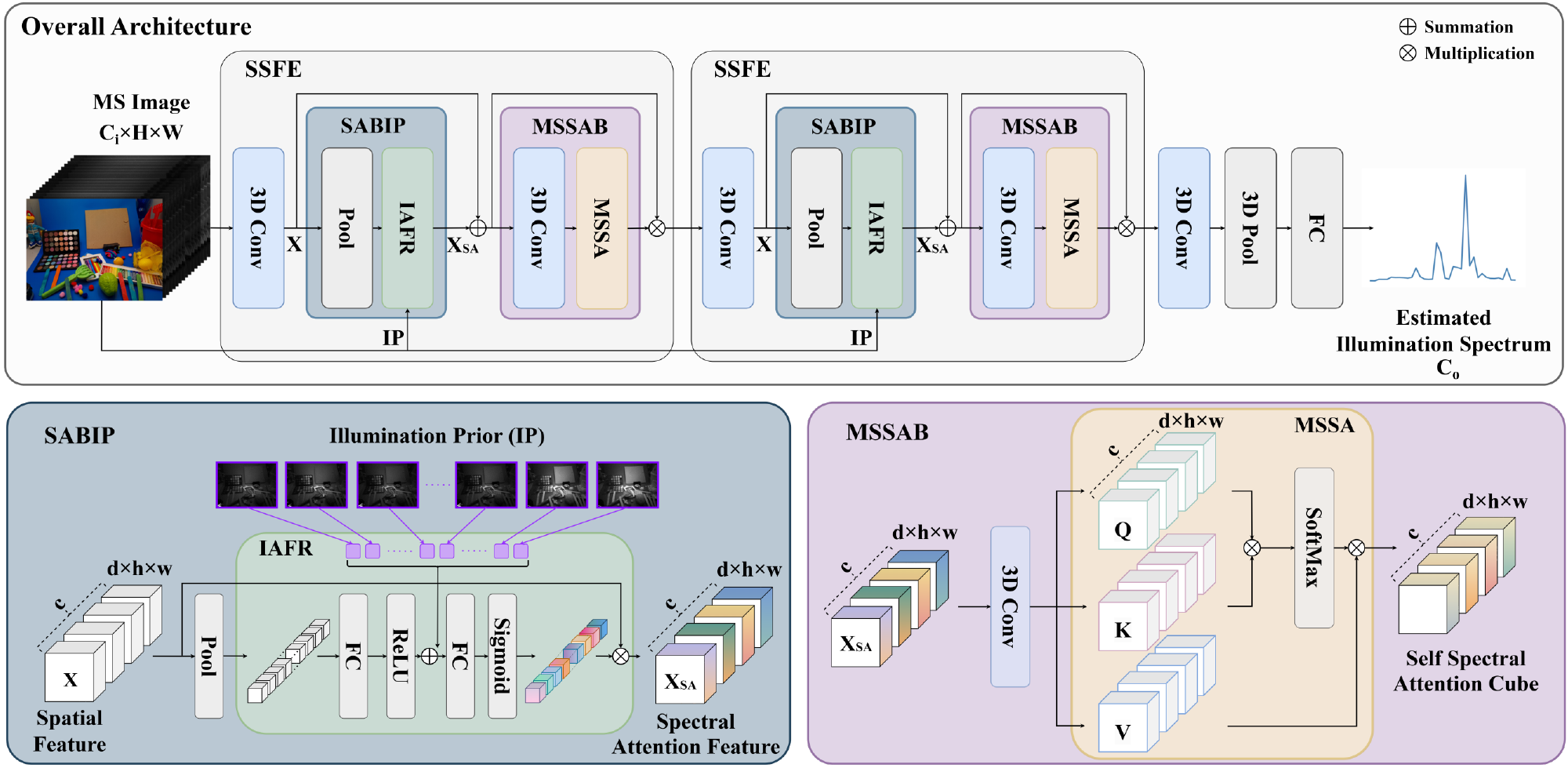}}
    \caption{Proposed architecture with two successive SSFEs.}
    \label{Overall architecture}
\end{figure*}

\subsection{Problem Formulation}

An image captured by a sensor under the assumption of Lambertian reflectance can be described as: 
\begin{equation}\label{eq:im_formulation}
\mathbf{I}_{k}(x,y)=\int_{\Lambda} \mathbf{L}(\lambda )\mathbf{R}(x,y,\lambda )\mathbf{C}_{k}(\lambda )d\lambda , k \in [1 \cdots N],
\end{equation}
where $\mathbf{I}_{k}(x,y)$ represents an MS image sample at a coordinate $(x,y)$ for the $k$-th channel, and $\mathbf{L}(\lambda )$ is an illuminant SPD. Note that $\mathbf{L}(x,y,\lambda )$ is approximated to $\mathbf{L(\lambda )}$ assuming the uniform illuminant within a single image. $\mathbf{R}(x,y,\lambda )$ and $\mathbf{C}_{k}(\lambda )$ represent the surface spectral reflectance and sensor spectral sensitivity, respectively. $k$ is the channel index, which ranges from 1 to $N$, where $N$ is the total number of the channels. $\lambda $ and $\Lambda$ denote wavelength and the wavelength range of the electromagnetic spectrum, respectively. 

By estimating the illumination spectrum accurately, the effects of lighting can be removed from the image, which would enable the recovery of the surface reflectance and the corresponding intrinsic color characteristics of the objects in the image. Due to the inherent spectral entanglement between illumination and surface reflectance, estimating the illuminant spectrum accurately is particularly challenging~\cite{tominaga1989SVD}.

\subsection{Overall Architecture}

\cref{Overall architecture} illustrates the overall architecture of the proposed network.
This network produces an illuminant spectrum vector $\mathbf{L} \in \mathbb{R}^{{C_{o}}}$ that approximates $\mathbf{L(\lambda )}$, from an input image $\mathbf{I} \in \mathbb{R}^{{C_{i}}\times H \times W}$, where $H$ and $W$ are the height and width, respectively, and $C_i$ and $C_o$ are the input and output channel sizes, respectively. These terms are described in \cref{ssec:Experiment_Setting}.

As shown in \cref{Overall architecture}, the proposed network comprises two successive spatio-spectral feature extractors (SSFEs). A 3D convolution is performed to generate spectral features, which are processed using two spectral attention modules. The first module contains a SABIP, which exploits the illuminant prior (IP) and focuses on  the most important spectral channels. The second module includes an MSSAB, which captures useful spectral information available across the spectral channels.

\subsection{Spatio-Spectral Feature Extractor}

An SSFE generates a spatio-spectral feature from an input MS feature. The proposed network contains two successive SSFEs, each of which comprises three key components\textemdash3D convolution, SABIP, and MSSAB\textemdash tailored to extract useful features from an MS image for effective ISE.

Upon entering the SSFE, the input feature goes through a 3D convolution ~\cite{3dconv} and is delivered to the SABIP. Unlike traditional 2D convolution kernels, 3D convolution kernels move over both the spatial and channel dimensions, thus ensuring that spectral details are not overlooked during convolutional operations.

\subsubsection{Spectral Attention Block with Illuminant Prior}

The SABIP is introduced to exploit the spectral correlation within the channels of an input spectral feature. It produces a spectral attention cube feature $\mathbf{X_{SA}}\in \mathbb{R}^{c\times d\times h\times w}$ (\cref{Overall architecture}) using a channel-wise 3D feature recalibration method based on an IP vector. Here, $c$ represents the number of feature channels, $d$ represents the spectral resolution, and h and w (height and width) represent the spatial dimensions (height and width) respectively.
The IP can be obtained using the following equation:
\begin{equation}
\begin{aligned}\label{eq:illuminant_prior}
\mathbf{IP}_{k} = \frac{\int \mathbf{I}_{k}(x,y)dx dy }{\int dx dy}, k \in [1 \cdots N]
\end{aligned}
\end{equation} where $\mathbf{IP}_{k}$ represents the IP for the $k$-th channel, which ranges from 1 to $N$. The IP is computed by normalizing the sum of all pixels in each channel of an input MS image with respect to the total number of pixels. 

\begin{figure}[!t]
    \centerline{\includegraphics[width=0.95\columnwidth]{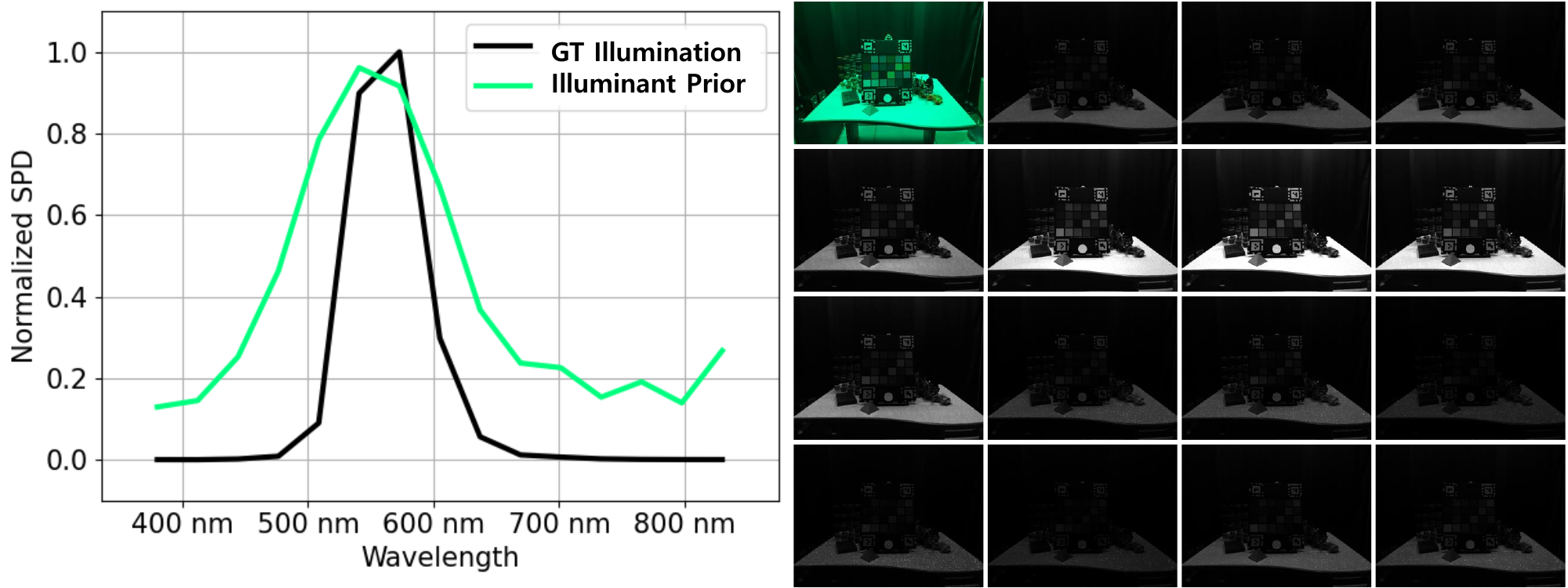}}
    \caption{Comparison between illuminant spectrum and IP vector for each channel of MS image. Left: Comparison between GT illuminant spectrum and IP vector. Right: RGB-rendered image (top left) and each channel of MS image.}
    \label{Im:IlluminantPrior} 
\end{figure}

As described in the image formation model in \cref{eq:im_formulation}, the response of each MS channel is jointly determined by the illuminant SPD, the surface reflectance, and the sensor spectral sensitivity. Therefore, the IP vector, computed as the channel-wise spatial mean response, is interpreted as an image-derived sensor-weighted illumination cue. Under the global-illumination condition and the gray-world assumption that the average reflectance of a sufficiently diverse scene is relatively achromatic, the IP vector can provide coarse information about the sensor-weighted illuminant distribution. As shown in \cref{Im:IlluminantPrior}, the IP vector is correlated with the overall spectral tendency of the scene illumination. Thus, we use the IP vector as a physical prior that provides coarse spectral guidance to the attention module for ISE.


\cref{Overall architecture} illustrates the implementation of the SABIP, which first aggregates global information through the spatial average pooling of the input spatial feature $\mathbf{x}\in \mathbb{R}^{c\times d\times h\times w}$. This process preserves spectral information and creates an embedding of a feature in $\mathbb{R}^{(c\times d) \times 1\times 1}$, where $c$ and $d$ represent spectral information. Then, this spectral embedding feature goes through an illuminant-aware feature recalibrator (IAFR), which uses a fully connected (FC) layer with the ReLU activation function to compute the interdependency within the spectral embedding feature. To the intermediate feature, we then add an IP vector approximating a scene illuminant that directs the network to focus on the illuminant-relevant spectral characteristics, as follows:
\begin{equation}
   \mathbf{x'} = \mathrm{ReLU}(\mathrm{FC}_{1}(\mathbf{x}))+\mathbf{IP},
\end{equation} where $\mathbf{x'}$ is the IP-added intermediate representation. This representation is further refined by being passed through another FC layer and a sigmoid activation, which produces a vector containing crucial spectral information related to the illuminant:
\begin{equation}
   \mathbf{s} = \sigma(\mathrm{FC}_{2}(( \mathbf{x'}))),
\end{equation} where $\mathbf{s}$ is the illuminant-related spectral vector.

Finally, $\mathbf{s}$ is then multiplied with the input feature map to perform spectral recalibration, which generates a spectrally recalibrated feature ($\mathbf{X_{SA}}$):
\begin{equation}
   \mathbf{X_{SA}} = \mathbf{x} \odot \mathbf{s},
\end{equation} where $\odot$ denotes the Hadamard product. This process enhances spectral representations by incorporating illuminant-related information into them.

\subsubsection{Multihead Spectral Self Attention Block}
The MSSAB captures meaningful inter-channel relationships and identifies the most important channels in $\mathbf{X_{SA}}$. This is achieved based on the MSSA scheme from an existing transformer model~\cite{vaswani2017attention}.
However, while \cite{vaswani2017attention} focused on spatial resolution, our spectral self-attention scheme uses embeddings for spectral dimensions. It calculates the similarities between the channels within $\mathbf{X_{SA}}$ to determine their relative importance. This process includes a 3D convolution to preserve spectral information from feature. The query, key, and value\textemdash denoted as $\mathbf{Q}$, $\mathbf{K}$, and $\mathbf{V}$\textemdash are obtained through a 3D convolution on $\mathbf{X_{SA}}$
\begin{equation}
\begin{aligned}\label{KQV}
\mathbf{Q} = \mathbf{K} = \mathbf{V} = \text{Conv3D}(\mathbf{X_{SA}}),
\end{aligned}
\end{equation}
where the dimensions of each embedding are $\mathbf{Q}, \mathbf{K}, \mathbf{V} \in \mathbb{R}^{c \times d \times h\times w}$.
These embeddings are then divided into the designated number of heads, which enables the calculation of attention scores, before being passed to the spectral self-attention block. 
Through its multi-head attention manner, MSSAB exploits channel dependencies from the perspectives of different heads, allowing the network to implicitly learn subtler information. 

By leveraging this two-stage attention processing involving the SABIP and MSSAB, the SSFE analyzes spectral correlations in a progressively refined manner. The SABIP first recalibrates the features using global illumination context, and the MSSAB then builds upon these enhanced features to learn consistent spectral relationships across the image. This attention mechanism enables the network to learn meaningful global spectral relationships across the image, focusing on the inherent correlations between different wavelengths reflecting the illumination in the scene.

\subsection{Objective Function}

When estimating an illuminant spectrum, it is crucial to capture the relative differences across spectral channels, which encode the illuminant’s chromatic characteristics. In our formulation, the orientation of the illuminant spectrum vector corresponds to its relative spectral shape (chromaticity), whereas the magnitude is associated with intensity-dependent photometric factors (illuminance) that can vary with exposure and acquisition conditions. Therefore, we employ angular error (AE) as the loss function, which measures agreement in the spectral shape while being invariant to global magnitude differences. Specifically, AE is defined as
\begin{equation}\label{eq:AE_definition}
   \mathcal{L}_{\text{AE}} = \text{AE}(\hat{\mathbf{L}},\mathbf{L}) = \arccos{\left(\frac{\hat{\mathbf{L}} \cdot \mathbf{L}}{\|\hat{\mathbf{L}}\| \|\mathbf{L}\|}\right)},
\end{equation}
where $\mathbf{L}$ and $\hat{\mathbf{L}}$ are the GT and predicted illumination spectrum vectors, respectively.

\subsection{Spectral-Domain Transform} \label{sec:cross-domain}
\subsubsection{Transformation Methods} 
We propose a spectral-domain transformation method that utilizes sensor-agnostic high-dimensional illuminant estimates from a deep learning model. The model is trained using high-dimensional GT lighting information acquired using a spectroradiometer to accommodate various spectral configurations. We use two specific spaces: the MS camera domain, which can be transformed using a camera sensitivity function (CSF), and the XYZ tristimulus space which can be transformed using a color-matching function (CMF).

The transformation from a high-dimensional illuminant spectrum to the low-dimensional MS camera domain can be expressed as
\begin{equation} \label{eq:transform_MS}
    \mathbf{\hat{L}}_{\mathrm{MS}}^{l} = \mathbf{M}_{\mathrm{CSF}}    \hat{\mathbf{L}}_{\mathrm{MS}}^{h}, 
\end{equation}
where $\hat{\mathbf{L}}_{\mathrm{MS}}^{h}$ and $\mathbf{\hat{L}}_{\mathrm{MS}}^{l}$ represent the estimated high-dimensional and low-dimensional illuminant spectra in MS camera domain, respectively, and $\mathbf{M}_{\mathrm{CSF}}$ represents a transformation matrix for the CSF (\cref{im:sensor_response_curve}).

For the trichromatic color space, specifically the XYZ color space, we use the CIE-1931 CMF~\cite{cie1931} as a transformation matrix to convert the illuminant into XYZ tristimulus values, which can be expressed as
\begin{equation}\label{eq:transform_XYZ}
    \mathbf{\hat{L}}_{\mathrm{XYZ}} = \mathbf{M}_{\mathrm{CMF}}   \hat{\mathbf{L}}_{\mathrm{MS}}^{h}, 
\end{equation} where $\mathbf{\hat{L}}_{\mathrm{XYZ}}$ is the illumination color vector in the XYZ color space and $\mathbf{M}_{\mathrm{CMF}}$ is the transformation matrix for the CIE-1931 CMF.

Through the aforementioned spectral-domain transformation, a single training process in the high-dimensional MS space allows the model to be applied across various domains without additional training. This ensures that the model performs consistently across different imaging domains, without being affected by specific sensor characteristics. Hence, our approach is highly practical for real-world applications.

\subsubsection{Mathematical Analysis of Domain Transformation} \label{subsec:SVD}

We mathematically analyze the high-dimensional illumination estimation and its evaluation across diverse lower-dimensional domains, focusing on which components of the high-dimensional estimated SPD affect the lower-dimensional target-domain response. Let $\mathbf{M}_{\mathrm{tar}} \in \mathbb{R}^{K\times N}$ denote a target-domain transform matrix from an $N$-dimensional MS spectral domain to a $K$-dimensional target color or sensor domain, where $K<N$. Here, $\mathbf{M}_{\mathrm{tar}}$ can correspond to either a CMF-based transformation for the CIE XYZ space or a CSF-based transformation for a target camera sensor space. Because angular error is computed from the normalized inner product between the ground-truth and estimated illuminants, we analyze the projected-domain inner product to identify which high-dimensional spectral components affect the target-domain evaluation. Since both the GT and estimated high-dimensional SPDs are projected by the same matrix, the inner product in the lower-dimensional target domain can be written as follows:
\begin{equation} \label{eq:cmf_svd}
    \begin{split}
    ({\mathbf{L}_{\mathrm{MS}}^{l}})^{\mathrm{T}} (\mathbf{\hat{L}}_{\mathrm{MS}}^{l})  &= (\mathbf{M}_{\mathrm{tar}} \mathbf{L}_{\mathrm{MS}}^{h})^{\mathrm{T}} (\mathbf{M}_{\mathrm{tar}}\mathbf{\hat{L}}_{\mathrm{MS}}^{h})\\
    &= ({\mathbf{L}_{\mathrm{MS}}^{h}})^{\mathrm{T}} ({\mathbf{M}_{\mathrm{tar}}}^{\mathrm{T}} \mathbf{M}_{\mathrm{tar}}) (\mathbf{\hat{L}}_{\mathrm{MS}}^{h})\\
    &= ({\mathbf{L}_{\mathrm{MS}}^{h}})^{\mathrm{T}} \sum_{i=1}^{r} \left(\lambda_{i}  \mathbf{q}_{i}\mathbf{q}_{i}^{\mathrm{T}} \right) \mathbf{\hat{L}}_{\mathrm{MS}}^{h} \\
    &= \sum_{i=1}^{r} \lambda_{i}  (({\mathbf{L}_{\mathrm{MS}}^{h}})^{\mathrm{T}}\mathbf{q}_{i}) (\mathbf{q}_{i}^{\mathrm{T}} \mathbf{\hat{L}}_{\mathrm{MS}}^{h}),
    \end{split}
\end{equation}
where $\mathbf{L}_{\mathrm{MS}}^{l} \in \mathbb{R}^{K}$ and $\mathbf{L}_{\mathrm{MS}}^{h} \in \mathbb{R}^{N}$ denote the target-domain illuminant representation and the high-dimensional illuminant SPD, respectively. The $\lambda_i$ and $\mathbf{q}_i$ are non-zero eigenvalues and corresponding eigenvectors of 
${\mathbf{M}_{\mathrm{tar}}}^{\mathrm{T}}\mathbf{M}_{\mathrm{tar}}$, where the rank $r=\operatorname{rank}({\mathbf{M}_{\mathrm{tar}}}^{\mathrm{T}}\mathbf{M}_{\mathrm{tar}})=\operatorname{rank}(\mathbf{M}_{\mathrm{tar}})\leq K$. Since ${\mathbf{M}_{\mathrm{tar}}}^{\mathrm{T}} \mathbf{M}_{\mathrm{tar}}$ is symmetric positive semi-definite, only these $r$ non-zero eigen-directions contribute to the projected-domain inner product. Therefore, when an $N$-channel high-dimensional spectrum is transformed into a $K$-channel target representation, at most $K$ spectral eigen-directions contribute to the converted representation. Components of the high-dimensional SPD lying in the null space of $\mathbf{M}_{\mathrm{tar}}$ do not affect the transformed low-dimensional representation.

Consequently, the transformation can be interpreted as selecting the target-observable spectral directions from the high-dimensional SPD. The non-zero eigen-directions, $\mathbf{q}_{1}, \mathbf{q}_{2}, \ldots, \mathbf{q}_{r}$ with $r\leq K$, determine the converted target-domain representation, whereas the remaining directions, $\mathbf{q}_{r+1}, \ldots, \mathbf{q}_{N}$, lie in the null space of the target transformation matrix and do not affect the converted result. 
Thus, the target transformation matrix does not arbitrarily reduce the estimated high-dimensional SPD. Instead, it retains the spectral components observable in the target space. 
A CMF-based matrix retains components relevant to the human visual XYZ space, whereas a CSF-based matrix retains components relevant to a target camera sensor space. This enables training-free transformation of the estimated illuminant SPD to different target color or sensor spaces using physically defined transformation matrices.
\section{Multispectral Image Dataset}
\label{sec:Multispectral image datset}
\begin{figure}[h!]
    \centerline{\includegraphics[width=\columnwidth]{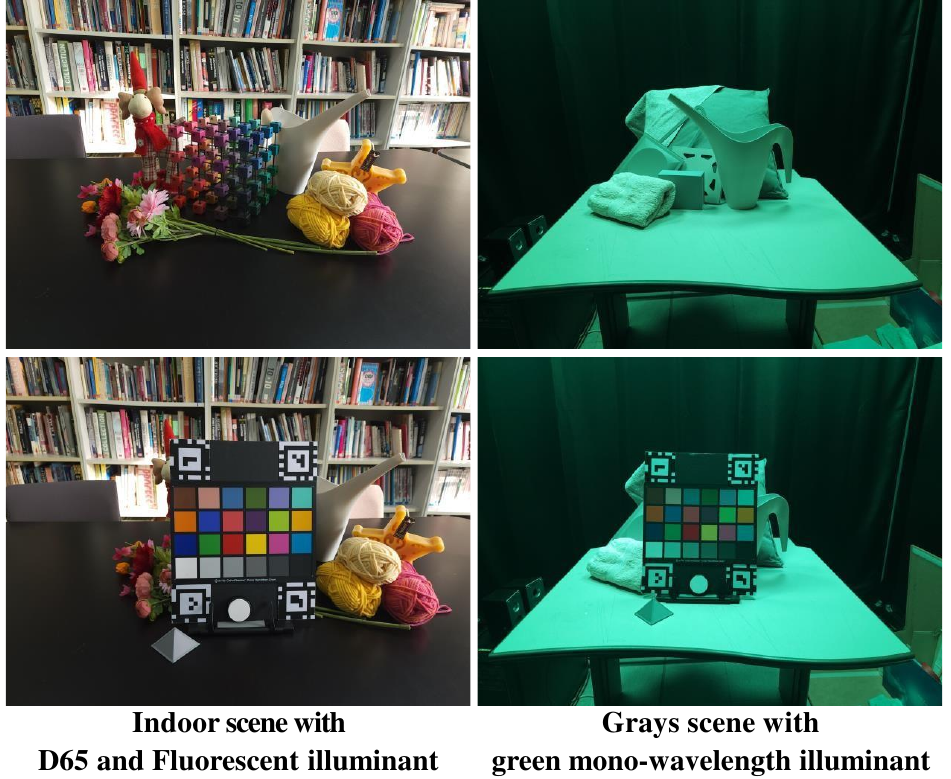}}
    \caption{Examples of images from MILD dataset. Top: Scenes without references. Bottom: Same scenes with references.} \vspace{-3mm}
    \label{MSI_dataset_sample}
\end{figure}

We propose the Multispectral Image dataset with Lighting Diversity (MILD) dataset, which comprises 15-channel MS images captured under various illumination conditions, ranging from mono-wavelength light sources to indoor illuminants such as D65 and fluorescent lighting. Thus, the MILD dataset can facilitate research on illumination estimation across diverse lighting environments on MS domains. The MILD dataset comprises 66 standard indoor scenes with D65 and fluorescent lighting, as well as 1405 laboratory scenes with diverse artificial lighting containing 42 mono-wavelength illuminants. \cref{MSI_dataset_sample} shows the sample images of the MILD dataset.

In the following sections, we detail the acquisition process for our dataset, including the MS image sensors used, the lighting conditions, and the GT illuminant measurement procedure.

\subsection{Multispectral Image Sensor}
\label{sub_sec:MS_senser}
\begin{figure}[h!]
    \centerline{\includegraphics[width=0.9\columnwidth]{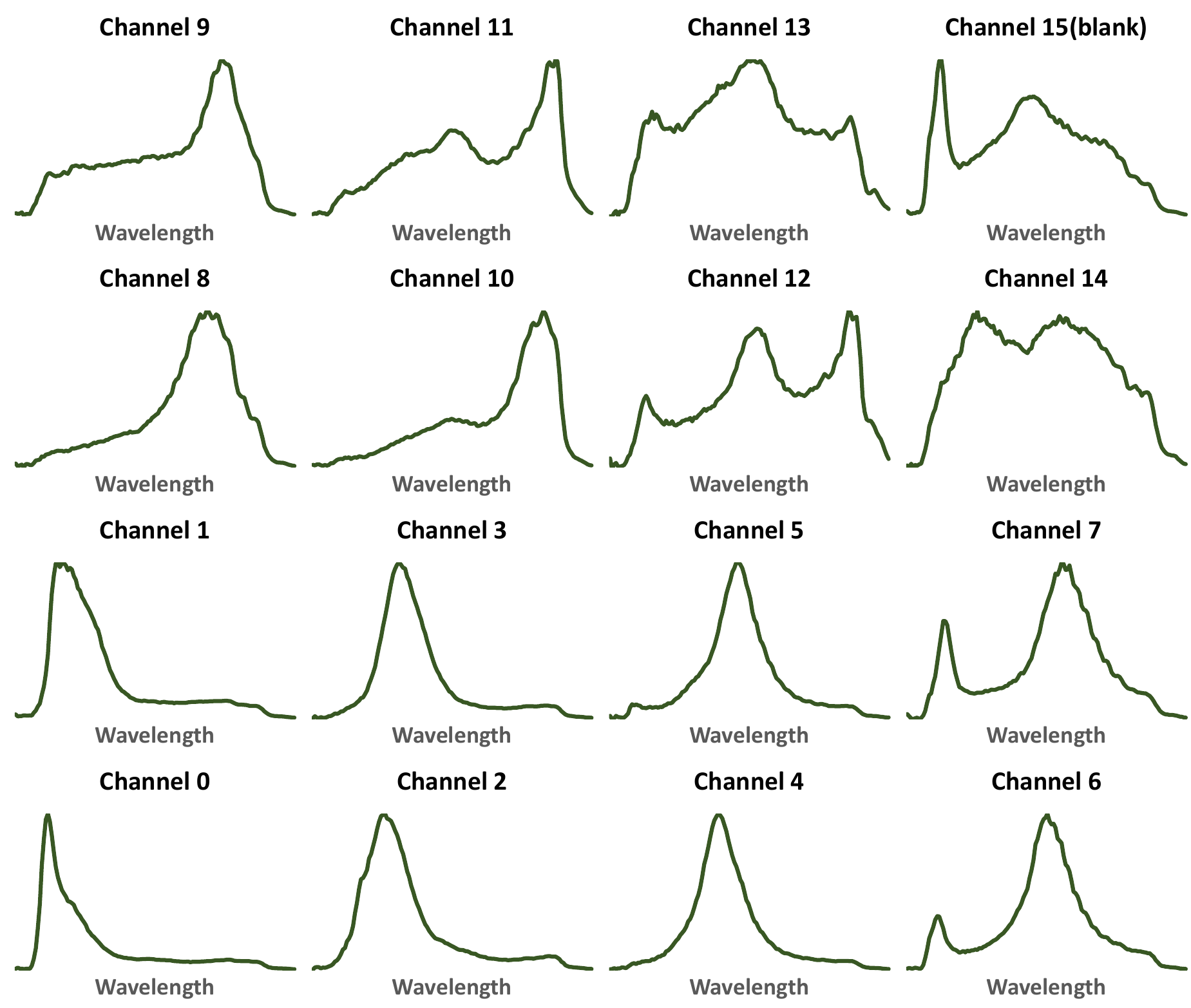}}
    \caption{Normalized spectral response for each channel of MS image sensor, with the spectrum normalized to the range [0,1].}
    \label{im:sensor_response_curve}
\end{figure}

The MS images are obtained using MS image sensors comprising 16 distinct channels, ranging from 380 \textit{nm} to 835 \textit{nm}. A Fabry-Perot type filter is used as a spectral bandpass filter, the spectral transmission band of which is set by adjusting the reflector and cavity length. This filter replaces the R, G, B color filter on a pixel of a CIS for a smartphone, of which the raw image resolution is 2584$\times$1936. Instead of a Bayer array, our spectral filter forms a 4$\times$4 filter array. \cref{im:sensor_response_curve} shows the spectral response curve of each channel according to the location of each channel pixel in the 4$\times$4 array. 
The channels are numbered 0\textendash 14 in ascending order of their peak wavelengths, with Channel 15 being a blank channel, which does not have any spectral filter. Since this blank channel does not provide specific spectral information, unlike the other 15 channels, it offers limited value for illumination estimation tasks. Therefore, the MILD dataset utilizes only the first 15 channels for such tasks. The images acquired by the MS image sensor undergo minimal processing, which includes the following steps: First, the black level in the raw data is removed, and the bad pixel is corrected using the median filter in the same spectral channel. Next, the lens shading is corrected using a precalculated gain map, whereafter demosaicing is performed to obtain each channel image by sampling pixels of the same channel across all the 4$\times$4 arrays; hence, each channel image has a resolution of 646$\times$484 pixels. Finally, noise reduction is performed on each channel image using a bi-lateral filter \cite{tomasi1998bilateral}.
 \vspace{-1mm}

\vspace{-2mm}
\subsection{Lighting Conditions}
\begin{figure}[h!]
    \centerline{\includegraphics[width=\columnwidth]{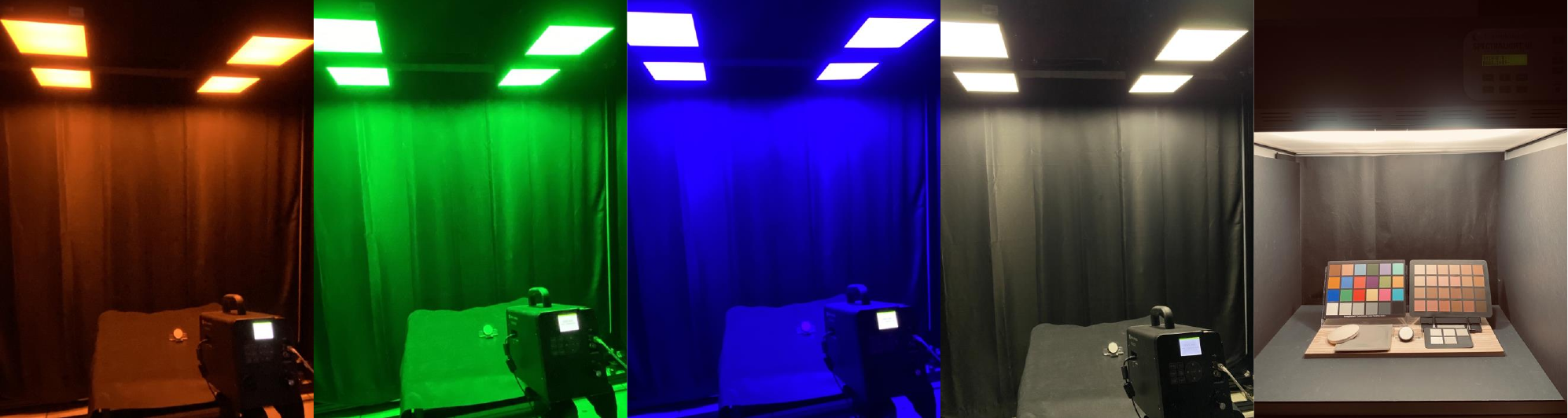}}
    \caption{MILD dataset acquisition environment.}
    \vspace{-3mm}
    \label{supp_im:dataset_env1}
\end{figure}

The MILD dataset encompasses two distinct indoor lighting conditions. The images contain either multi-color objects or only gray objects subjected to various illuminants. The first lighting condition represents common indoor environments illuminated by standard light sources, such as D65 or fluorescent lighting. The second lighting condition corresponds to highly controlled scenarios in a lighting booth (LED CUBE) powered by 15 channels of LED used to stimulate of the various illuminants, including mono-wavelength colors. The mono-wavelength illuminants contain red, green, blue, and yellow lighting, which can be considered extreme lighting. \cref{supp_im:dataset_env1} presents an example of the dataset acquisition environment established for various light sources. Our dataset comprises images captured under highly controlled conditions. To prevent any influence of unintended light sources, the images were acquired in a darkroom. 

\begin{figure}[h!]
    \centerline{\includegraphics[width=1\columnwidth]{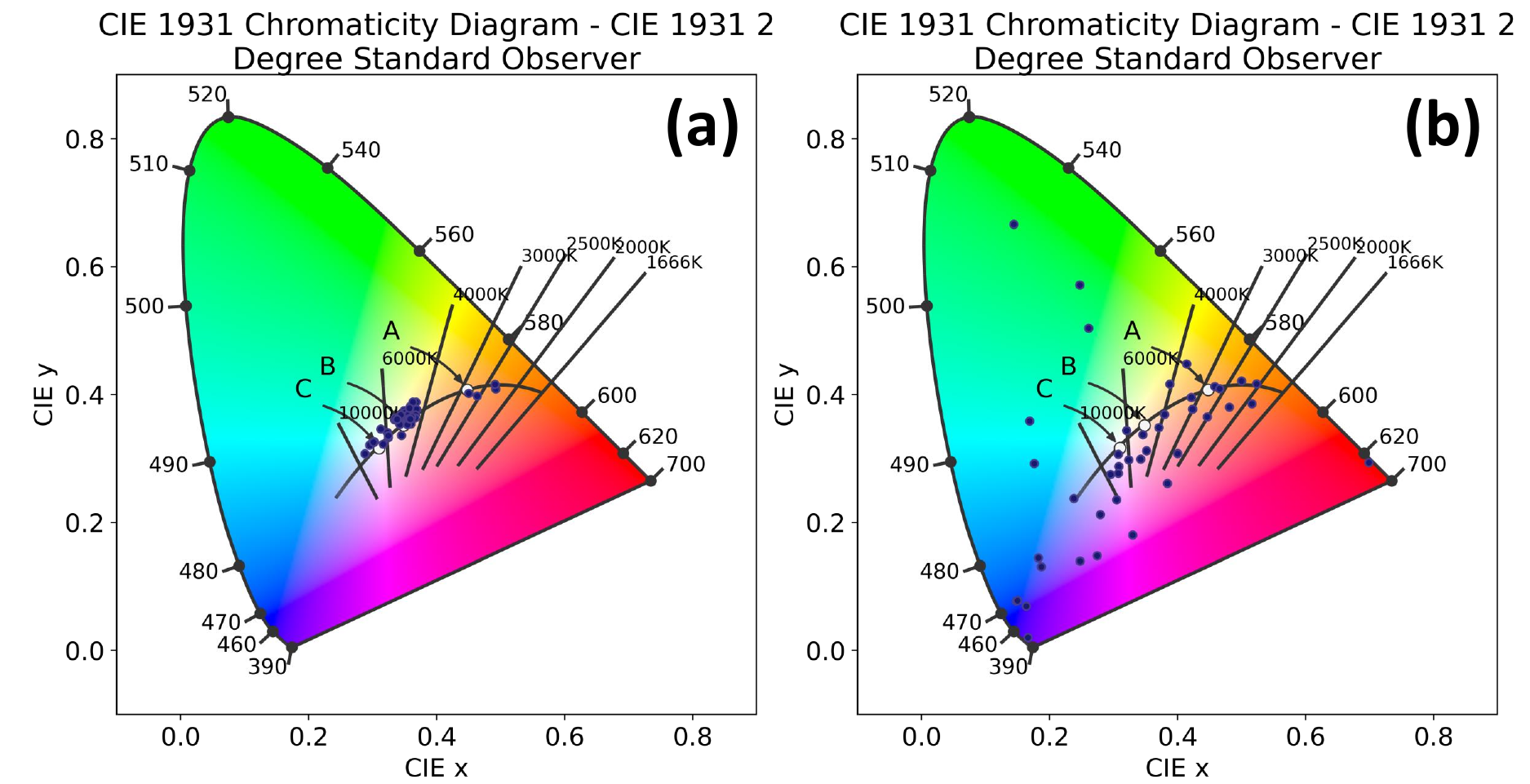}}
    \caption{CIE-xy chromaticity coordinates of illuminants in MILD dataset: (a) Illuminants that are not mono-wavelength. (b) Illuminants with mono-wavelength spectrum.}
    \label{supp_im:cct_check}
\end{figure}

The aforementioned extreme mono-wavelength conditions represent corner cases in illumination estimation situations where the illuminant’s SPD is concentrated at a single wavelength; thus, such cases are particularly challenging to handle for existing benchmark methods. 
\cref{supp_im:cct_check} illustrates the mapping of the illuminants in our MS dataset to the Planckian locus. As shown in \cref{supp_im:cct_check} (a), illuminants such as D65 and fluorescent light lie on the Planckian locus, representing standard lighting conditions typical in natural and artificial indoor environments. In contrast, mono-wavelength illuminants (\cref{supp_im:cct_check}(b)) deviate substantially from this locus. Under such mono-wavelength lighting conditions, properly rendering the intrinsic colors of objects becomes challenging; hence, these cases are particularly difficult for illumination estimation compared with standard lighting conditions. 
Therefore, the inclusion of monochromatic lighting is crucial for verifying the robustness of illumination estimation algorithms, and it also aligns with the recent trend of color constancy studies related to aesthetic and emotional responses \cite{hurlbert2019challenges}.

\vspace{-5mm}
\subsection{Acquisition of Ground-Truth Illumination}
\begin{figure}[h!]
\centerline{\includegraphics[width=\columnwidth]{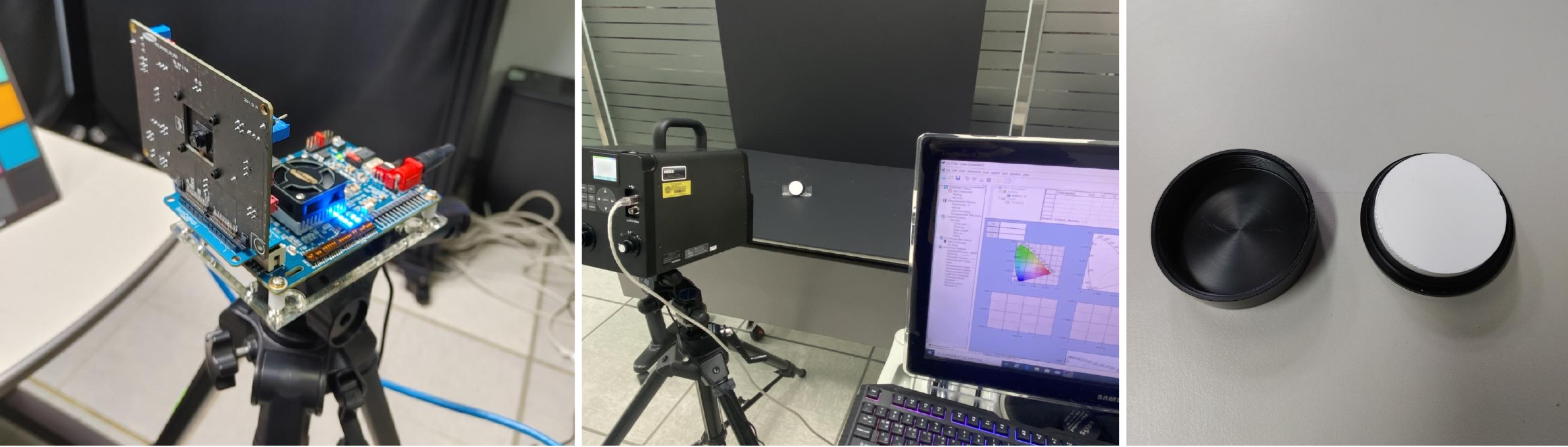}}
    \caption{Equipment used for MILD dataset acquisition. Left: MS image sensor. Middle: Spectroradiometer. Right: White reference.} \vspace{-4mm}
    \label{supp_im:dataset_env2}
\end{figure}

Every image in the MILD dataset contains references for GT illumination measurement as the Macbeth Color Checker and the KRISS White Standard along with a gray-patched object. The 3D gray-patched reference has four sides with mid-tone grays (N5\textendash N6) to enable the detection of illuminants from different directions. The GT spectrum of each illuminant, reflected from the KRISS White Standard reference, is measured using a spectroradiometer (CS-2000, Minolta) at 1 \textit{nm} intervals. These illuminant spectra are resampled to 36 channels spanning from 380 \textit{nm} to 730 \textit{nm} at 10 \textit{nm} intervals. \cref{supp_im:dataset_env2} shows the MS sensor used to capture the images, the spectroradiometer used to acquire the GT illuminant spectra, and the KRISS White Standard reference.

\begin{table*}[t!]
\centering
\caption{Performance comparison between proposed method and existing methods on real-world MS dataset (BeyondRGB).}
\resizebox{0.9\textwidth}{!}{%
\begin{tabular}{c|ccccc|ccccc}
\toprule[1.5pt]
\multirow{3}{*}{Method}  & \multicolumn{5}{c|}{Lab} & \multicolumn{5}{c}{Field} \\ \cmidrule{2-11} 
& \multicolumn{5}{c|}{$\Delta \mathrm{A_{MS}} \downarrow $} & \multicolumn{5}{c}{$\Delta \mathrm{A_{MS}} \downarrow $}  \\
 & mean & 25\% & median &75\% & std & mean & 25\% & median &75\% & std \\ \midrule
  WP w/ FFNN ~\cite{erba2024rgb} &11.26  &6.13 &9.39 &15.28 &6.35  &8.86 &4.74 &8.43 &9.91 &5.47 \\
  GW w/ FFNN ~\cite{erba2024rgb} &10.08 &5.72 &8.51 & 14.90&5.45 & 8.32 &5.08 &6.90 &9.98 &4.60 \\
 GE$^\text{1st}$ w/ FFNN ~\cite{erba2024rgb} &10.87 & 6.50 &9.34 &14.19 &5.81 &8.68 &5.00 &8.43 &11.98 &4.71 \\ 
 PWIR~\cite{pixelwise} &9.70 &5.57 &8.11 &12.24  &5.84  & 9.32 &5.13 &8.72 &12.08 & 4.96 \\
 BeyondRGB~\cite{glatt2024beyond}  & 5.92  & 4.04 &5.39 &8.01 & 2.92  & 7.22 &3.31  &6.14 &9.89 & 5.54  \\
 Ours  &\textbf{2.51}  & \textbf{1.27} & \textbf{1.88}  &\textbf{3.18} &\textbf{1.80}  &\textbf{4.92}  &\textbf{2.80} &\textbf{4.53} &\textbf{5.70} &\textbf{2.83}\\ 
\bottomrule[1.5pt]

\end{tabular}%
}
\label{Tab:BeyondRGB_HS}
\end{table*}
\section{Experimental Results}
\label{sec:Experiment}

We validate our method across complementary evaluation settings, including comprehensive benchmarks and challenging scenarios, as follows:
(i) Datasets with \emph{spectroradiometer-measured} illuminant spectra (spectroradiometer GT): BeyondRGB (\cref{subsec:EX_BeyondRGB}) and MILD (\cref{sub_sec:EX_MILD}).
(ii) Robustness under mono-wavelength illumination using the MILD(m) subset.
(iii) Reflectance benchmarks (KAUST/CAVE) evaluated under \emph{synthesized} illuminants (\cref{sub_sec:EX_KAUST_CAVE}).
(iv) \emph{Training-free} spectral-domain transformation to MS/XYZ target spaces, demonstrating cross-sensor generalization without retraining (\cref{sub_sec:EX_cross-domain}).
(v) Ablation studies quantifying the contribution of each component (\cref{sub_sec:EX_ablation}).
(vi) Extensibility to intensity-level and multi-source illumination 
estimation via physical attenuation modeling (\cref{sub_sec:EX_extensibility}).
(vii) White-balancing application demonstrating the practical 
utility of the estimated illuminant spectra (\cref{sub_sec:EX_WB}).

\subsection{Experimental Settings}\label{ssec:Experiment_Setting}
\subsubsection{Dataset}
We evaluate our proposed method on various spectral datasets. For spectral image evaluation, we employ the BeyondRGB~\cite{glatt2024beyond} and MILD datasets, which include both reflectance and illumination information. The BeyondRGB dataset comprises 1680 diverse scenes, including 1208 collected in lab settings and 472 in real-world field environments. Each scene is captured using a 16-channel MS sensor covering 380\textendash 730 \textit{nm} wavelengths, and the GT illuminant spectrum is measured using a spectroradiometer at 10 \textit{nm} intervals across the same spectral range; thus, an illuminant spectrum vector with 36 channels is produced.

The MILD dataset comprises indoor images capturing diverse objects under various lighting conditions, with each image consisting of 15 input MS channels covering 380\textendash 835 \textit{nm}. The GT illuminant spectrum is measured in the same manner as for BeyondRGB. We also create the MILD(m) dataset, which contains 42 images with only mono-wavelength illumination captured in laboratory settings, selected from among the 1,471 images in the MILD dataset. The mono-wavelength illuminations contain red, green, blue and yellow lights composed of single or double spectral peaks.

For spectral reflectance evaluation, we utilize the KAUST ~\cite{deepunrollingmultispectral} and CAVE~\cite{yasuma2010generalized} datasets, which contain only reflectance information without illuminant components. 
From these datasets, we select 31 spectral channels, spanning wavelengths of 400\textendash 700 \textit{nm} with 10 \textit{nm} sampling intervals. To simulate realistic illumination conditions, we generate spectral images by multiplying 59 standard illuminants ~\cite{mansencal_2022_7367239} with the surface reflectance of each pixel in 409 and 32 reflectance images from the KAUST and CAVE datasets, respectively.

\subsubsection{Training and Testing}
For experimental purposes, we use the data-split protocol from ~\cite{glatt2024beyond} for BeyondRGB and the protocol from ~\cite{deepunrollingmultispectral} for KAUST and CAVE. The CAVE dataset is employed exclusively as a testing benchmark to assess model generalization.

The input dimensions of the network are determined by the specifications of the camera sensor used to capture the images in each dataset. For BeyondRGB and MILD, while the input channels ($C_{in}$) correspond to the respectively original image channels, the number of output channels ($C_{out}$) is increased to 36 to match the GT illuminant spectrum measured by the spectroradiometer. The input and output channels for KAUST and CAVE match the respective dataset's original image channels. 

We utilize AE, a widely adopted metric in ISE research, as our evaluation metric. It is calculated using \cref{eq:AE_definition}:
\begin{equation}
    \Delta \mathrm{A} = \text{AE}(\hat{\mathbf{L}},\mathbf{L}),
\end{equation} where ${\mathbf{L}}$ and $\hat{\mathbf{L}}$ represent the GT and estimated illuminant spectra, respectively.

In the following subsections, $\Delta \mathrm{A_{MS}}$ represents the AE in the MS space. For the BeyondRGB and MILD datasets, since the GT illuminant spectrum is acquired with a 36 dimensional vector, $\Delta \mathrm{A_{MS}}$ is computed over 36 output channels. For the KAUST and CAVE datasets, it is computed over 31 channels. Additionally, mean-$\Delta \mathrm{A_{MS}}$ and std-$\Delta \mathrm{A_{MS}}$ denote the average and standard deviation of the AEs, respectively.

\begin{figure*}[h!]
    \centerline{\includegraphics[width=\textwidth]{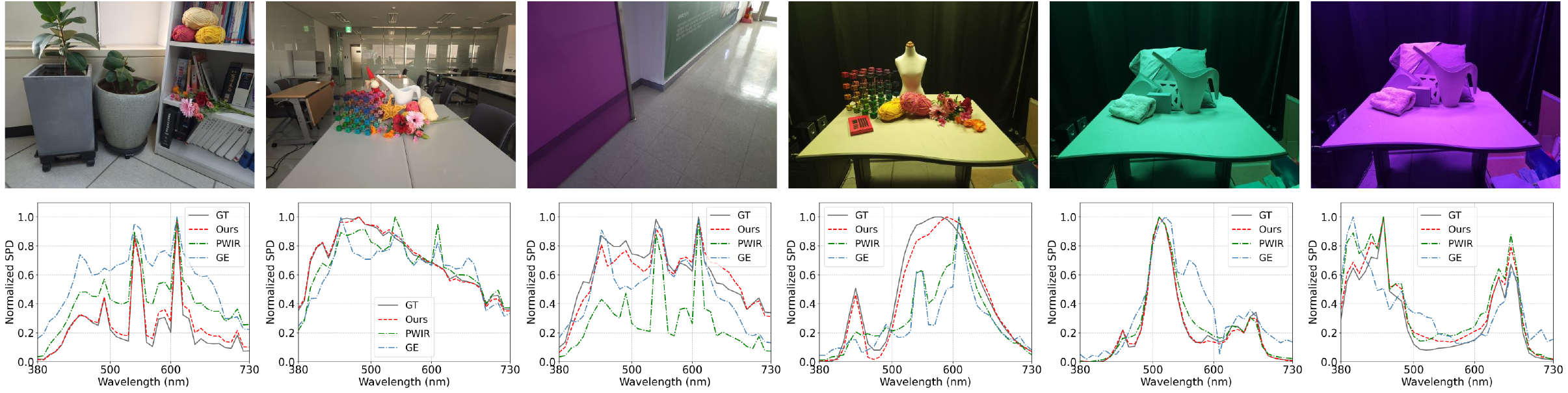}}
    \caption{Qualitative estimations of illuminant compared across methods on MILD dataset. Top row: RGB-rendered input MS image. Bottom row: SPD of estimated illuminant.} \vspace{-1.5mm}
    \label{fig:qual_results}
\end{figure*}

 \vspace{-0.5mm}
\subsection{BeyondRGB Dataset}
\label{subsec:EX_BeyondRGB}

The BeyondRGB dataset is composed of ``Lab'' and ``Field'' subsets, which respectively include images of various objects with diverse colors and textures in indoor scenes and real-world images captured in outdoor scenes. To evaluate the performance of our method, we compare it with both statistical and learning-based methods. The statistics-based methods include versions of widely adopted illumination estimation algorithms extended to cover MS imaging ~\cite{khan2017illuminant}, White Point(WP)~\cite{land1985retinex}, GW~\cite{grey-world}, and first order GE(GE$^\text{1st}$) ~\cite{grey-edge} enhanced with a feed-forward neural network ~\cite{erba2024rgb}.
Among deep learning-based methods, we employ PWIR~\cite{pixelwise} and BeyondRGB~\cite{glatt2024beyond} because other approaches require reflectance data for training ~\cite{deepunrollingmultispectral} or require the input dimension to match the number of spectral bands in the GT spectra ~\cite{zheng2015illumination,ISNL}.

\cref{Tab:BeyondRGB_HS} lists the accuracies of the tested methods, where the 25\%, median, and 75\% in \cref{Tab:BeyondRGB_HS} indicate 25th, 50th, and 75th percentiles of the AE, respectively. The results demonstrate that our proposed method significantly outperforms existing approaches. On the Lab and Field datasets, its AEs are 2.51 and 4.92, and its mean-$\Delta \mathrm{A_{MS}}$ is substantially lower than that of the SOTA method ~\cite{glatt2024beyond}, specifically by 57.6\% and 31.9\%, respectively. The proposed method also exhibits a narrow interquartile range (IQR; Lab: 1.27\textendash 3.18, Field: 2.80\textendash 5.70) and the lowest standard deviation (Lab: 1.80, Field: 2.83), which demonstrates its exceptional consistency. Specifically, for Field images depicting complex real-world scenes with more diverse objects, our method's IQR (2.80\textendash 5.70) compared with the second-best method's IQR (3.31\textendash 9.89) further highlights the superiority of our approach in handling complex real-world scenes. This low error distribution underscores the reliability and robustness of our method.

\subsection{MILD Dataset}
\label{sub_sec:EX_MILD}
\cref{Tab:MILD_EX} compares the accuracy of the proposed method with that of the benchmark methods on the MILD dataset. To provide a broader comparison, we include RGB-based and spectral image-based illuminant estimation methods. Since RGB-based methods cannot estimate a 36-channel illuminant SPD, their $\Delta \mathrm{A}_{MS}$ values are marked as N/A, and all methods are additionally compared in the CIE XYZ space using $\Delta \mathrm{A}_{XYZ}$.
Our method exhibits the highest accuracy on the proposed datasets, i.e., both MILD and MILD(m). It achieves average AE of 3.18$^{\circ}$ and 7.44$^{\circ}$ with standard deviations of 3.69 and 2.95 on MILD and MILD(m), respectively, indicating stable illumination estimation performance. Compared with BeyondRGB, the proposed method reduces the mean $\Delta \mathrm{A}_{MS}$ from 4.96$^{\circ}$ to 3.18$^{\circ}$ on MILD and from 11.49$^{\circ}$ to 7.44$^{\circ}$ on MILD(m). In the XYZ-space comparison, our method also achieves the lowest mean $\Delta \mathrm{A}_{XYZ}$ among both RGB and MS baselines, with 1.34$^{\circ}$ on MILD and 5.09$^{\circ}$ on MILD(m). The results verify that the proposed method delivers remarkable performance compared with the benchmark methods, not only under ordinary lighting conditions but also in extreme conditions.

\begin{table}[h!]
\centering
\caption{Performance comparison between proposed method and existing methods on MILD dataset.}
\resizebox{0.95\columnwidth}{!}{%
\begin{tabular}{cccccc}
\toprule[1.5pt]
 \multirow{2}{*}{Dataset} & \multirow{2}{*}{Method} & \multicolumn{2}{c}{$\Delta \mathrm{A_{MS}} \downarrow $} & \multicolumn{2}{c}{$\Delta \mathrm{A_{XYZ}} \downarrow $} \\ \cmidrule{3-6} 
  & & mean & std & mean &std \\ \midrule
  \multirow{9}{*}{MILD}
  &FC4 ~\cite{FC4} & N/A & N/A & 1.94 & 2.32 \\
  &C4 ~\cite{C42020} & N/A &N/A  & 1.84 &2.05 \\
  &TLCC ~\cite{tang2022transfer} &N/A  &N/A  &3.63  &2.59  \\
  &WP w/ FFNN ~\cite{erba2024rgb} &15.62&8.25 & 7.33  &5.47\\
  &GW w/ FFNN ~\cite{erba2024rgb} &12.90 &6.22 & 5.28 &4.58\\
  &GE$^\text{1st}$ w/ FFNN ~\cite{erba2024rgb} & 13.68 & 7.82 &5.00  &4.38\\ 
  &PWIR~\cite{pixelwise} &9.56  & 5.34  &7.20  &5.33\\
  &BeyondRGB ~\cite{glatt2024beyond}  &4.96  &4.49  &2.19  &3.10  \\
  &Ours  &\textbf{3.18}  &\textbf{3.69}   &\textbf{1.34}  &\textbf{1.93}\\ 
 \midrule \midrule
  \multirow{9}{*}{MILD(m)}
  &FC4 ~\cite{FC4} & N/A & N/A & 5.86 & 5.27 \\
  &C4 ~\cite{C42020} & N/A &N/A  &6.20 &3.81\\
  &TLCC ~\cite{tang2022transfer} &N/A  &N/A  &5.88  &\textbf{3.28}\\
  &WP w/ FFNN ~\cite{erba2024rgb}&23.21 &12.73 &14.07 &5.26\\
  &GW w/ FFNN ~\cite{erba2024rgb} &19.15 &8.86 &8.87 &5.36  \\
  &GE$^\text{1st}$ w/ FFNN ~\cite{erba2024rgb} &21.13 &9.85 &7.89  &5.41\\ 
  &PWIR~\cite{pixelwise} &13.46  &3.16  &15.99  &8.69\\
  &BeyondRGB ~\cite{glatt2024beyond} &11.49 &7.26	&7.50 & 8.00\\
  &Ours  &\textbf{7.44}  &\textbf{2.95}  &\textbf{5.09}  &3.59 \\ 
 \bottomrule[1.5pt]
\end{tabular}%
} \vspace{-2mm}
\label{Tab:MILD_EX}
\end{table}


\cref{fig:qual_results} compares the estimated illuminant spectrum and the corresponding rendered RGB image from the MILD dataset between the proposed method and the benchmark methods. Based on the curves of the estimated illuminant spectrum for the tested methods, our model successfully estimates the GT, even in mono-wavelength image samples.

\subsection{KAUST and CAVE Datasets}
\label{sub_sec:EX_KAUST_CAVE}

We comparatively evaluate the proposed method against existing ISE methods, namely the following statistics-based and learning-based methods: GE \cite{khan2017illuminant}, LRMF \cite{zheng2015illumination}, ISNL \cite{ISNL}, PWIR \cite{pixelwise}, DUN \cite{deepunrollingmultispectral} and BeyondRGB ~\cite{glatt2024beyond}.

Table \ref{Tab:KAUST} compares the performance of these methods on the KAUST and CAVE datasets. The proposed method achieves a mean-$\Delta \mathrm{A_{MS}}$ of 4.32$^{\circ}$ on the KAUST dataset, showing a 40.8\% improvement over the SOTA, BeyondRGB. On the CAVE dataset, the proposed method achieves a mean-$\Delta \mathrm{A_{MS}}$ of 4.66$^{\circ}$, demonstrating a 29.8\% improvement compared with BeyondRGB.
Overall, the learning-based methods, such as DUN, BeyondRGB, and ours, which are designed for MS images, outperform the statistics-based models, such as GE, LRMF, and ISNL. Our method also displays stable illuminant estimation performance with standard deviations of 2.17 and 1.95 on the KAUST and CAVE datasets, respectively. Notably, our approach maintains robust performance even on the CAVE reflectance data, which were unseen during the training phase.
These results demonstrate that our method not only achieves superior accuracy relative to existing algorithms but also exhibits stable performance on unseen data.

\begin{table}[h!] 
\centering
\caption{Performance comparisons between proposed method and existing methods on KAUST and CAVE datasets.}
\resizebox{0.95\columnwidth}{!}{%
\begin{tabular}{cccc}
\toprule[1.5pt]
Dataset & Method & mean-$\Delta \mathrm{A_{MS}} \downarrow $ & std-$\Delta \mathrm{A_{MS}} \downarrow $ \\ \midrule
\multirow{7}{*}{KAUST}
 &GrayEdge~\cite{khan2017illuminant}  & 11.46 & 8.02  \\
 &LRMF~\cite{zheng2015illumination} & 24.06 & 23.49  \\
 &ISNL~\cite{ISNL} & 20.05 & 11.46   \\
 &PWIR~\cite{pixelwise} & 28.07 & 21.20   \\
 &DUN~\cite{deepunrollingmultispectral} & 9.17 & 6.30 \\
 &BeyondRGB~\cite{glatt2024beyond} & 7.30  & 4.45  \\
 &Ours  &\textbf{4.32}  &\textbf{2.17} \\ 
\midrule  \midrule
\multirow{7}{*}{CAVE}
 &GrayEdge~\cite{khan2017illuminant}  &18.33  &14.89 \\
 &LRMF~\cite{zheng2015illumination}  &20.05  &15.47  \\
 &ISNL~\cite{ISNL}  &22.92  &16.62  \\
 &PWIR~\cite{pixelwise}  &34.95  &21.20  \\
 &DUN~\cite{deepunrollingmultispectral}   &16.04  &13.75  \\
 &BeyondRGB~\cite{glatt2024beyond}   &6.64  &3.62  \\
 &Ours  &\textbf{4.66}  &\textbf{1.95}  \\
 \bottomrule[1.5pt]
\end{tabular}%
} \vspace{-2mm}
\label{Tab:KAUST}
\end{table}

\subsection{Spectral-Domain Transformation}
\label{sub_sec:EX_cross-domain}

We evaluate the proposed network across various channel configurations, comparing the transformation and end-to-end approaches. The transformation approach involves training a network to produce a high-dimensional illuminant vector and transform it to a low-dimensional sensor space without any retraining. Conversely, in the end-to-end approach, the output illuminant vector is directly produced from the output neurons. To achieve this, we define $C_{in}$ as the number of channels in the input image, $C_{out}$ as the number of channels in the output illumination, and $C_{target}$ as the number of channels in the evaluation space.

We evaluate cases where the network predicts a higher-dimensional illumination compared with the input image ($C_{in}<C_{out}$) as well as cases where it predicts illuminations with as many or fewer dimensions as the input image ($C_{in} \geq C_{out}$). This approach allows us to verify the neural network's ability to learn and estimate higher-dimensional spectral information even from lower-dimensional input.

For evaluation domains, designated as $C_{target}$, we focus on cases where the number of dimensions for the output illumination exceeds or equals that for the target domain ($C_{out} \geq C_{target}$), to validate the effectiveness of the transformation approach. This technique enables us to examine whether performance is maintained when converting high-dimensional illumination estimations to various spaces.

The AE in each domain is calculated as follows:
\begin{equation}
\begin{gathered}
\Delta \mathrm{A}_{\mathrm{MS}}^{l} = \mathrm{AE}(\mathbf{\hat{L}}_{\mathrm{MS}}^{l},\mathbf{L}_{\mathrm{MS}}^{l}), \\
\Delta \mathrm{A_{XYZ}} = \mathrm{AE}(\mathbf{\hat{L}_{XYZ}},\mathbf{L_{XYZ}}), 
\end{gathered}
\end{equation} where $\Delta \mathrm{A}_{\mathrm{MS}}^{l}$ and $\Delta \mathrm{A_{XYZ}}$ represent the AE in lower-dimensional MS camera space and the XYZ space, respectively. Each illumination in the MS space and XYZ space is transformed as described in \cref{sec:cross-domain}.

For the cross-domain evaluation, we use the proposed model as the backbone network, modifying only the input and output dimensions for different channel configurations, and test it on the BeyondRGB dataset, which includes information on camera spectral sensitivity.

\cref{tab:cross_domain_MS} shows the results for the MS camera space as the target space, with $C_{target} = 15$. The results show that estimating a 36-channel illuminant and then transforming it to a 15-channel illuminant using CSF matrix yields better results than direct end-to-end estimation of 15-channel illuminants; specifically achieving the lower average angular error (transformation: 1.05, end-to-end: 1.13). This suggests that high dimensional illuminant estimation using a deep learning network captures more comprehensive spectral information, which is preserved effectively during transformation to lower-dimensional space. Specifically, once a network is trained on the high-dimensional GT, it can be applied to the MS domain without being retrained.

\begin{table} [h!]
    \caption{Results of cross domain evaluation, where the target domain is a 15-channel camera sensor space.}
    \centering
    \begin{tabular}{cc|cc|c}
        \toprule[1.5pt]
         $C_{in}$ & $C_{out}$ & $C_{target}$ & Transformation & mean-$\Delta \mathrm{A}_{\mathrm{MS}}^{l} \downarrow $ \\ \midrule
         15&36 &15 &CSF & \textbf{1.05}\\
         15&15 &15 &End-to-End & 1.13\\ \bottomrule[1.5pt]
    \end{tabular}
    \label{tab:cross_domain_MS}
\end{table}

\cref{tab:cross_domain_XYZ} shows the results for the XYZ space as the target space, with $C_{target}=3$. We observe a similar trend, i.e., higher-dimensional illumination estimation captures richer spectral information, which is well-preserved even through transformation. Comparing rows 2 and 3, we observe that the illuminant estimation performance is better when using 15-channel images as input than when using 3-channel images for the same scene. This indicates that utilizing MS images with more spectral information can lead to better illuminant estimation performance, even when estimating illuminants in lower-dimensional spaces.

\begin{table}[h!]
    \caption{Results of cross domain evaluation, where the target domain is a 3-channel XYZ tristimulus space.}
    \centering
    \begin{tabular}{cc|cc|c}
        \toprule[1.5pt]
         $C_{in}$ & $C_{out}$ & $C_{target}$ & Transformation & mean-$\Delta \mathrm{A_{XYZ}} \downarrow $ \\ \midrule
         15&36 &3 &CMF &\textbf{1.25} \\
         15&3 &3 &End-to-End & 1.29 \\
         3&3 &3 &End-to-End & 1.35 \\ \bottomrule[1.5pt]         
    \end{tabular}
    \label{tab:cross_domain_XYZ}
\end{table}

\subsection{Complexity Analysis}
To examine the practical feasibility of the proposed framework, we analyze the computational complexity of the proposed high-dimensional illuminant SPD estimation. Since our framework estimates a 36-channel illuminant SPD and transforms it to lower-dimensional target spaces using a fixed CSF/CMF-based spectral-domain transformation, we compare two controlled variants: the proposed 15ch$\rightarrow$36ch model and a direct 15ch$\rightarrow$3ch XYZ prediction model. Both settings use the same 15-channel MS input, SSFE backbone, 3D convolution, and MSSAB modules. Therefore, this comparison directly evaluates the overhead introduced by predicting a 36-channel illuminant SPD instead of a 3-channel target illuminant. The complexity was measured with batch size 1 and $256\times256$ input resolution on a single NVIDIA RTX A6000 GPU.

\begin{table}[h!]
\caption{Complexity comparison between the proposed high-dimensional SPD estimation setting and direct XYZ prediction setting.}
\centering
\resizebox{0.85\columnwidth}{!}{%
\begin{tabular}{ccccc}
\toprule[1.5pt]
Model setting & Params & MACs & Memory & Time \\
\midrule
15ch$\rightarrow$36ch & 16.30M & 0.542G & 126.2MB & 3.38ms \\
15ch$\rightarrow$3ch  & 16.27M & 0.542G & 126.0MB & 3.31ms \\
\bottomrule[1.5pt]
\end{tabular}%
}
\label{Tab:complexity_analysis}
\end{table}

As shown in \cref{Tab:complexity_analysis}, the proposed 15ch$\rightarrow$36ch setting introduces negligible overhead compared with direct 15ch$\rightarrow$3ch XYZ prediction. The proposed model increases the parameter count by only 0.03M, from 16.27M to 16.30M, while the rounded MAC count remains the same at 0.542G. The peak GPU memory changes only from 126.0MB to 126.2MB, and the inference time increases by only 0.07ms. These MAC values correspond to approximately 1.084 GFLOPs when one MAC is counted as two floating-point operations. This small overhead is expected because both settings share the same SSFE backbone and differ mainly in the final regression head. In addition, the CSF/CMF-based spectral-domain transformation is only a fixed matrix-vector multiplication applied to the predicted illuminant vector. These results show that the proposed framework maintains practical computational complexity while enabling flexible target-domain transformation without additional training.

\vspace{-2mm}
\subsection{Ablation Study}
\label{sub_sec:EX_ablation}

\cref{Tab:ablation} presents the results of an ablation study conducted to examine the impact of different components of network architecture on the network's performance. The ablation study was conducted on the BeyondRGB dataset using the combined test samples from both the Lab and Field subsets. The results indicate the effectiveness of the 3D Conv, SABIP, and MSSAB components of the proposed method. When using 3D Conv instead of 2D Conv, we observe an improvement of more than 0.76$^{\circ}$ in the MS domain. Adding SABIP reduces the average AE from 3.42$^{\circ}$ to 3.23$^{\circ}$. When incorporating all three modules, the proposed model achieves the best accuracy, with an average AE of 3.18$^{\circ}$ and the lowest standard deviation among all models (2.38). This finding demonstrates that each component contributes to the overall performance and stability of our network architecture.

\begin{table}[htb!]
\caption{Ablation study on network architecture with different combinations of our modules using BeyondRGB dataset.}
\centering
\resizebox{0.8\columnwidth}{!}{
\begin{tabular}{ccccc}
\toprule[1.5pt]
\multirow{2}{*}{{3D Conv}} & \multirow{2}{*}{{SABIP}} & \multirow{2}{*}{{MSSAB}} & \multicolumn{2}{c}{$\Delta \mathrm{A_{MS}} \downarrow $} \\ \cmidrule{4-5}
 &  &  & mean & std \\ \midrule
\ding{55} & \ding{55} & \ding{55} & 4.18 & 2.86  \\
\checkmark & \ding{55} & \ding{55} & 3.42 & 2.50  \\
\checkmark & \checkmark & \ding{55} & 3.23 & 2.48  \\
\checkmark & \checkmark & \checkmark & \textbf{3.18} &\textbf{2.38}
\\ 
\bottomrule[1.5pt]
\end{tabular}}  \vspace{-2mm}
\label{Tab:ablation}
\end{table}

To further verify whether the image-derived IP provides useful illumination-related guidance for learning, we conduct an additional controlled ablation study by replacing the IP with input-independent alternatives. Specifically, we compare the original IP with random gaussian noise, a uniform constant vector, and an input-independent learnable spectral vector. The random and constant variants preserve the same number of learnable parameters as the original SABIP, while removing scene-dependent illumination information. The learnable vector variant tests whether an input-independent learnable vector can replace the image-derived IP with minimal additional capacity.

\begin{table}[htb!]
\caption{Ablation study for the illuminant prior in SABIP using BeyondRGB dataset.}
\centering
\resizebox{0.95\columnwidth}{!}{
\begin{tabular}{ccccc}
\toprule[1.5pt]
 \multirow{2}{*}{Method} & \multicolumn{2}{c}{$\Delta \mathrm{A}_{MS} \downarrow $} & \multirow{2}{*}{\makecell{Additional\\Params}} & \multirow{2}{*}{IP} \\ 
\cmidrule{2-3}
 & mean & std &  &  \\ 
\midrule
SABIP w/o IP (random) &3.40 &2.50 &0 &\ding{55} \\
SABIP w/o IP (constant) &3.70 &2.69 &0 &\ding{55} \\
SABIP w/o IP (learnable vector) &3.29 &2.49 &15 &\ding{55} \\
SABIP (ours)& \textbf{3.18} &\textbf{2.38}& - &\checkmark \\
\bottomrule[1.5pt]
\end{tabular}}
\label{Tab:ablation_IP}
\end{table}

As shown in \cref{Tab:ablation_IP}, replacing the input-dependent IP with random noise or a constant vector increases the mean $\Delta \mathrm{A}_{MS}$ from 3.18$^{\circ}$ to 3.40$^{\circ}$ and 3.70$^{\circ}$, respectively. Since these variants have the same number of learnable parameters as the original SABIP, the degradation indicates that the image-derived IP provides useful illumination-related information for learning. The input-independent learnable vector also performs slightly worse than the original IP-based SABIP, even though it introduces additional learnable parameters. This confirms that the scene-dependent physical prior provides useful illumination information that cannot be fully replaced by an input-independent learnable vector. In sum, these results indicate that the performance gain of SABIP is not merely due to increased model capacity.

\begin{figure*}[t!]
    \centering
    \includegraphics[width=0.9\textwidth]{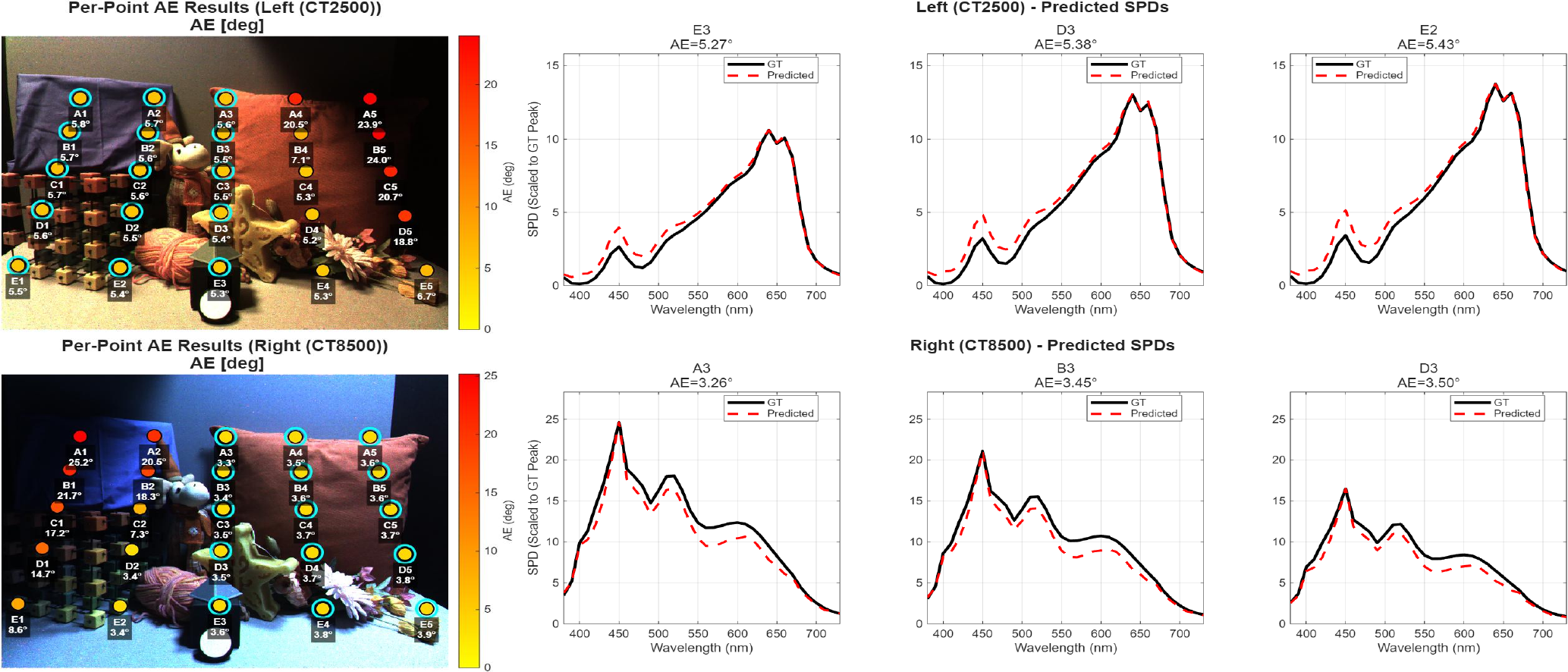}
    \caption{Single-source illumination estimation with the Yuksel attenuation model. The cyan circles indicate points directly illuminated by each lighting source, defined as spatially proximal to the source. The top row corresponds to the left-sided warm CT2500 illumination, and the bottom row corresponds to the right-sided cool CT8500 illumination.} 
    \label{fig:single_yuksel}
\end{figure*}

\subsection{Extensibility to Intensity-Level and Multi-Source Illumination Estimation.} \label{sub_sec:EX_extensibility} In this subsection, we conducted experiments demonstrating that our illumination estimation provides a sufficient basis not only for intensity-level modeling but also for multi-source illumination modeling when combined with a physically grounded attenuation model.

The measured illuminant SPD at a spatial location $(x,y)$ can be expressed as:
\begin{equation} \label{eq:single_source_spd}
F_\text{spd}(x,y) = I(x,y) \cdot \mathbf{S} = A \cdot f_a(d,r) \cdot \mathbf{S},
\end{equation}
where $ \mathbf{S} \in \mathbb{R}^{36}$ is the estimated scale-invariant spectral shape, $A$ is the absolute source intensity (a scalar constant), and $f_a(d,r)$ is the energy attenuation function based on the distance from the light source. Specifically, we leverage the point light attenuation model of Yuksel ~\cite{yuksel2020point}, which formulates spatially varying illumination intensity as
\begin{equation}
f_a(d,r)=\frac{2}{r^2}\left ( 1-\frac{d}{\sqrt{d^2+r^2}} \right ),
\end{equation}
where $d$ is the distance from the light source and $r$ is a virtual source radius.

In our framework, $\mathbf{S}$ is provided by the network output and treated as a scale-invariant spectral shape, since the network is trained using angular error. The absolute spatial intensity variation is therefore fitted separately using the magnitudes of the spectroradiometer-measured reference SPDs. Let $F_i^{\mathrm{ref}}$ denote the measured reference SPD at location $(x_i,y_i)$, and let $d_i=\sqrt{(x_i-x_s)^2+(y_i-y_s)^2+z_s^2}$ be the distance from the estimated light-source position $(x_s,y_s,z_s)$ to the $i$-th reference point. We estimate $(x_s,y_s,z_s)$ and $r$ by fitting the Yuksel attenuation function to the relative spatial fall-off of the measured SPD magnitudes over directly illuminated reference points $\mathcal{V}$ as
\begin{equation}
\scalebox{0.96}{$\displaystyle
\min_{x_s,y_s,z_s,r}
\sum_{i\in \mathcal{V}}
\left(\frac{f_a(d_i,r)}{\max_{k\in \mathcal{V}} f_a(d_k,r)}-\frac{\|F_i^{\mathrm{ref}}\|_2}{\max_{k\in \mathcal{V}}\|F_k^{\mathrm{ref}}\|_2}\right)^2.
$}
\end{equation}

After estimating the source geometry and $r$, the absolute intensity scale $A$ is computed as
\begin{equation}
A=\frac{\sum_{i\in \mathcal{V}}\|F_i^{\mathrm{ref}}\|_2}
{\sum_{i\in \mathcal{V}}f_a(d_i,r)}.
\end{equation}
Using the estimated $A$ and $r$ in \cref{eq:single_source_spd}, the spatially varying illuminant SPD can be recovered without additional network training.

For the multi-source case, this extends naturally via superposition
\begin{equation}F_\text{spd}(x,y) = A_1 \cdot f_a(d_1,r_1) \cdot \mathbf{S}_1 + A_2 \cdot f_a(d_2,r_2) \cdot \mathbf{S}_2.\label{eq:superposition}\end{equation}
where $\mathbf{S}_1$ and $\mathbf{S}_2$ are the spectral shapes of the two sources. 

In practice, the Yuksel model requires spatially distributed reference measurements or prior knowledge of the source geometry to define the source-to-surface distance.

\subsubsection{Single-source illumination intensity validation} To validate the physical modeling of illumination attenuation, we adapted the Yuksel model with our multispectral image data. We captured a scene under two independent single-source illuminants — a warm CT2500 source (peak ~640 nm) and a cool CT8500 source (peak ~450 nm) — and measured per-point absolute illuminant spectra using a spectroradiometer (Konica Minolta CS-2000) at 25 spatially distributed measurement locations. We then verified whether the Yuksel attenuation model can recover the per-point intensity distribution. As shown in \cref{fig:single_yuksel}, the fitted and measured illumination SPDs show strong agreement in direct illumination regions (indicated by cyan circles), where indirect effects such as cast shadows are excluded. As summarized in \cref{tab:single_yuksel}, the fitted illumination SPDs achieve mean AE of 5.57$^\circ$ for CT2500 and 3.62$^\circ$ for CT8500 in these regions, closely matching our network's output (reference source of Yuksel model) accuracy of 5.42$^\circ$ and 3.62$^\circ$, respectively. Furthermore, the fitted attenuation model achieves $R^2 > 0.97$ against the measured per-point GT illumination intensity in the direct illumination regions, where $R^2$ denotes the coefficient of determination between predicted and measured per-point intensity, confirming that the distance-dependent intensity variation is accurately captured by the physical model. While regions dominated by indirect illumination effects such as inter-reflections and cast shadows exhibit larger errors, as these secondary effects fall outside the scope of the physical attenuation model, the model achieves strong agreement in all directly illuminated regions, confirming its validity where the underlying physical assumptions hold.
These results validate that our spectral shape estimates, combined with the Yuksel physical model, can effectively capture intensity levels of light.

\begin{table} [h!]
    \caption{Single-source Yuksel model results. Reference (S) 
    denotes the angular error of the network's estimated 
    illuminant SPD used as the spectral shape input to 
    the Yuksel model, representing the baseline accuracy 
    before attenuation modeling.}
    \centering
    \begin{tabular}{c|c|cc}
        \toprule[1.5pt]
        Light Source & Region & \#pts & mean-$\Delta \mathrm{A_{MS}} \downarrow $ \\ 
        \midrule
        \multirow{4}{*}{CT2500} 
            & Reference (S) & 1 & 5.42° \\ \cmidrule{2-4}
            & Direct & 15 & 5.57° \\
            & Indirect      & 10 & 13.75° \\
            \cmidrule{2-4}
            & All           & 25 & 8.84° \\ 
        \midrule[1pt]
        \multirow{4}{*}{CT8500} 
            & Reference (S) & 1 & 3.62° \\  \cmidrule{2-4}

            & Direct   & 15 & 3.62° \\
            & Indirect      & 10 & 14.04° \\
            \cmidrule{2-4}
            & All           & 25 & 7.79° \\
        \bottomrule[1.5pt]
    \end{tabular}
    \label{tab:single_yuksel}
\end{table}

\begin{figure*}[t!]
    \centering
    \subfloat[Verification of the spectral superposition principle of linear combination of the two individually measured single-source spectra.\label{subfig:superposition}]{
        \includegraphics[width=0.9\textwidth]{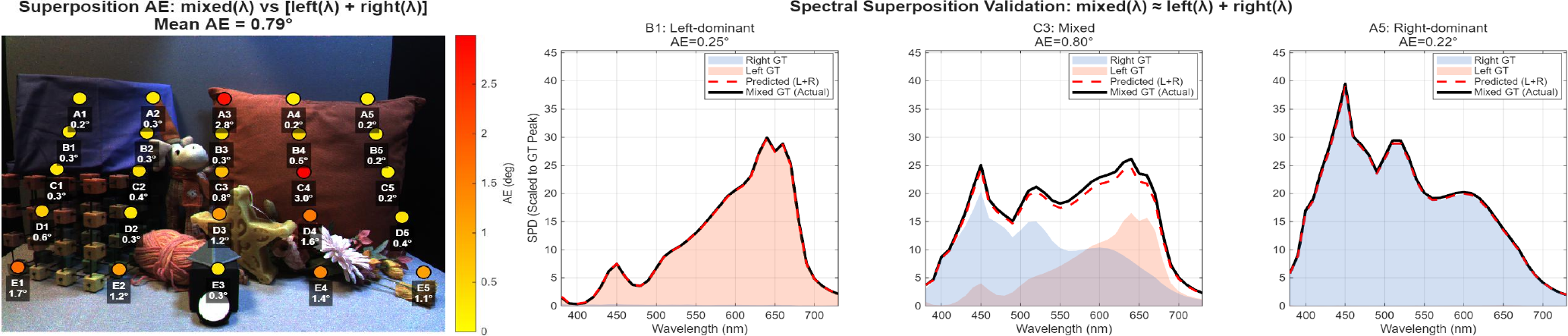}
    }
    
    \subfloat[Multi-source illumination estimation with the Yuksel attenuation model. The points indicated by cyan circles in the figures are those directly illuminated by both lighting sources.\label{subfig:mixed_yuksel}]{
        \includegraphics[width=0.9\textwidth]{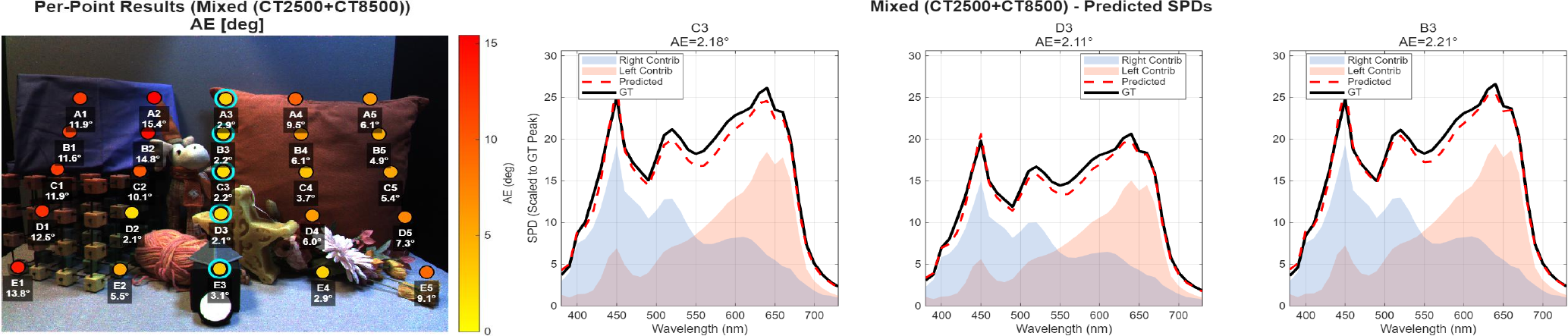}
    }
    \caption{Experimental results for spectral superposition verification and multi-source illumination estimation.}
    \label{fig:multi-source}
\end{figure*}

\begin{table} [h!]
    \caption{Multi-source Yuksel model results.}
    \centering
    \begin{tabular}{c|c|cc}
        \toprule[1.5pt]
         Light Source & Region & \#pts & mean-$\Delta \mathrm{A_{MS}} \downarrow $ \\ 
         \midrule
        \multirow{3}{*}{CT2500+CT8500} &Both Direct &5 &2.48$^\circ$  \\
        &Indirect & 20 & 8.54$^\circ$ \\          
        \cmidrule{2-4}
        &All & 25 & 7.33$^\circ$ \\
         \bottomrule[1.5pt]
    \end{tabular} \vspace{-2mm}
    \label{tab:mixed_yuksel}
\end{table}

\subsubsection{Multi-source illumination spectral superposition and intensity validation} To further validate the extensibility of our framework to multi-source scenarios, we conducted a mixed-illuminant experiment where both CT2500 and CT8500 sources were active simultaneously.

We first verified the spectral superposition principle, whether the mixed illuminant spectrum at each spatial point can be expressed as a linear combination of the two individually measured single-source spectra as expressed in ~\cref{eq:superposition}. Specifically, we compared the measured mixed-illuminant spectrum against the direct sum of per-point single-source spectroradiometer measurements at each of the 25 locations. Across all 25 measurement points, the mean angular error between the measured mixed spectrum and the predicted sum was 0.79$^\circ$, confirming that spectral superposition holds with high fidelity in this controlled setting.

Subsequently, we applied the Yuksel model independently to each source component and verified that the spatial intensity distribution of the multi-source scene can be recovered by superposing the two single-source Yuksel models using only the network-estimated spectral shapes $\mathbf{S}_1$ and $\mathbf{S}_2$ as reference inputs, as shown in \cref{subfig:mixed_yuksel}. As summarized in \cref{tab:mixed_yuksel}, in regions where both sources contribute direct illumination (5 points), the multi-source Yuksel model achieves a mean AE of 2.48$^\circ$, confirming that the mixed illumination estimation of two independently fitted Yuksel models holds with high fidelity. In contrast, points where one source is occluded (20 points) show larger errors (mean AE = 8.54$^\circ$), consistent with the single-source case, as indirect illumination effects dominate these regions and fall outside the scope of the physical model. Note that in the multi-source case, the AE inherently reflects the accuracy of spatially varying intensity ratios between spectrally distinct sources, since the mixed spectral shape at each location is determined by the relative intensity contributions of the two sources with different spectral characteristics. These results collectively demonstrate that our single-dominant illuminant estimation framework provides a physically grounded foundation for extending to more complex intensity and multi-illuminant scenarios.

\subsection{White-Balancing Result}\label{sub_sec:EX_WB}
We performed white-balancing (WB) on images from the MILD dataset using the estimated illuminants from the network compared in \cref{sec:Experiment}; the GT images are color-corrected with respect to the GT illuminants. As the input images and the estimated illuminants are both in the MS domain, we transform them into the RGB color domain for visualizing the WB results. Therefore, all images are rendered from the MS to the RGB domain. 
For MILD, we follow the DNG-based rendering pipeline provided with the BeyondRGB dataset, since MILD was captured using the same multispectral sensor configuration. The estimated illuminant SPD is first converted to CIE XYZ using the CIE 1931 color matching functions, and the resulting illuminant XYZ is mapped to the camera reference sensor space, and channel-wise WB gains are computed. The WB-corrected linear RAW image is then rendered to sRGB.

\begin{figure*}[h!]
    \centering
    \includegraphics[width=0.95\textwidth]{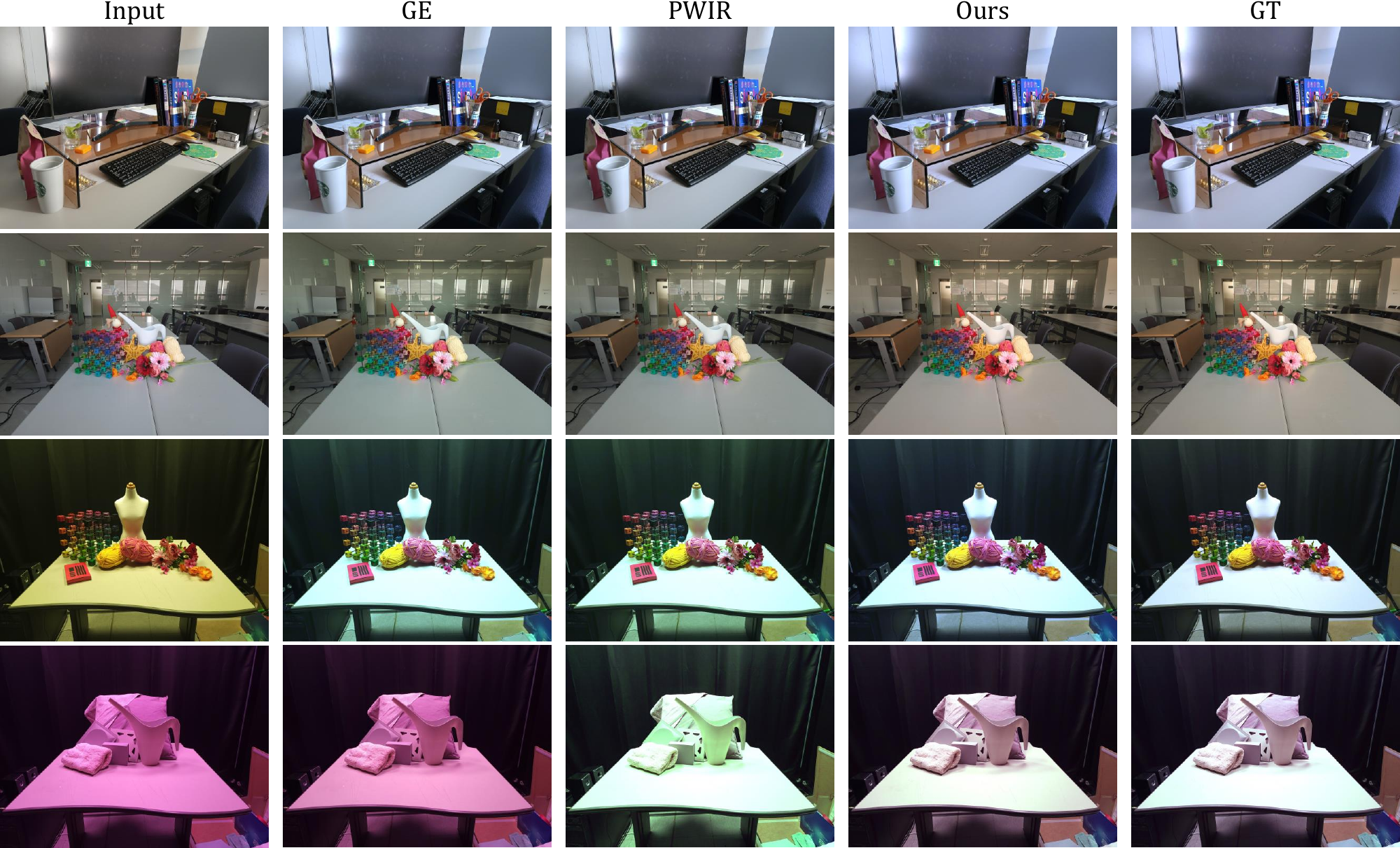}
    \caption{Results of WB on MILD dataset. From left to right: Original input image (1st); WB images obtained using illuminants estimated by GE (2nd), PWIR (3rd), and proposed model (4th); WB image obtained using GT illuminant (5th).}
    \label{fig:AWB_MILD}
\end{figure*}

The closer the color appearance is to the GT image, the better is the illuminant estimation, thus, the more effective is the WB.
\cref{fig:AWB_MILD} displays WB results on the MILD dataset with the estimated and the GT illuminants. WB results on other datasets can be found in the supplementary material.

\section{Discussion and Future Work}
\label{sec:D&F}
Our framework focuses on estimating the spectral power distribution of a single dominant illuminant from a multispectral image. This addresses the fundamental requirement of illumination estimations, which is to recover the spectral shape of the underlying light source. This is essential for subsequent tasks such as color correction, white balance and spectral rendering.
As demonstrated in ~\cref{sub_sec:EX_extensibility}, the estimated spectral shape serves as a sufficient basis for modeling intensity-level variation via physical attenuation models ($R^2 > 0.97$ for single-source intensity fitting) and for multi-source illumination estimation via spectral superposition (mean AE of 2.48$^\circ$ in directly illuminated regions). However, the current extension assumes known or estimable source positions, does not account for indirect illumination such as inter-reflections or cast shadows, and has been validated only in controlled indoor settings with neutral reference surfaces. 
The extension of this approach to unconstrained environments, for instance by integrating neural source detection, learning-based occlusion-aware attenuation, and end-to-end spatially-varying illuminant estimation directly within the learning-based network architecture, is a promising direction for future work.

\section{Conclusion}
\label{sec:Conclusion}
We developed an efficient deep learning-based illumination estimation model for MS images. The proposed SSFE block uses SABIP and MSSAB, which are designed to exploit rich spectral information effectively. By incorporating spectral attention mechanisms and IP, the proposed method optimally enhances spectral correlation and preserves illuminant-relevant spatial features.

To facilitate evaluation, we created the MILD MS dataset captured under diverse lighting conditions, which contains high-dimensional GT illumination spectra measured using spectroradiometer. This dataset can facilitate further research due to its higher spectral resolutions, which is crucial for deep models. While existing methods show limitations under extreme conditions, such as mono-wavelength lighting, the proposed model efficiently overcame such problems. Based on the experimental results, the proposed method outperformed the existing methods, especially under challenging lighting conditions. Furthermore, our approach leverages spectral-domain transformation, the effectiveness of which was validated both mathematically and experimentally. Additionally, we showed that the estimated single-dominant illumination SPD provides a physically grounded basis for extending to intensity-level and multi-source illumination modeling via physical attenuation models.

{
    \small
    \bibliographystyle{ieeenat_fullname}
    \bibliography{main}
}

\begin{IEEEbiography}[{\includegraphics[width=1in,height=1.25in,clip,keepaspectratio]{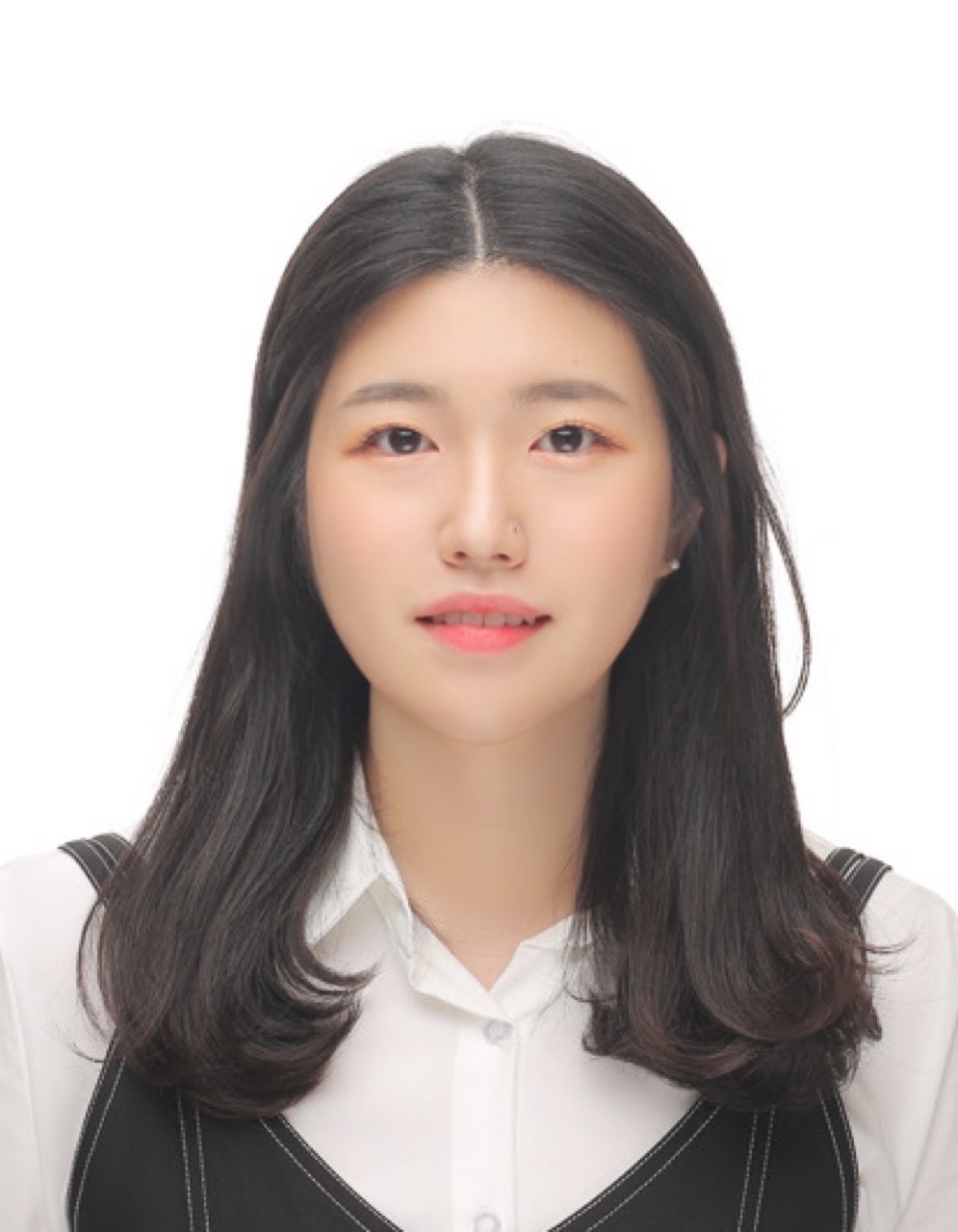}}] {Hyejin Oh} received the B.S. degree in electronics engineering from Ewha Womans University, Seoul, South Korea, in 2021, where she is currently pursuing the Ph.D. degree with the Department of Electronic and Electrical Engineering. In 2024, she was a Visiting Student with the Department of Mechanical and Industrial Engineering, University of Toronto, Toronto, ON, Canada, as part of the AI Excellence Global Innovative Leader Education Program. Her research interests include deep learning-based image and video understanding, computational imaging, multispectral image analysis, and immersive media compression. She has also participated in MPEG standardization activities related to immersive media compression. She was a co-recipient of the 2025 Sadaoki Furui Prize Paper Award from APSIPA Transactions on Signal and Information Processing.
\end{IEEEbiography}

\begin{IEEEbiography}[{\includegraphics[width=1in,height=1.25in,clip,keepaspectratio]{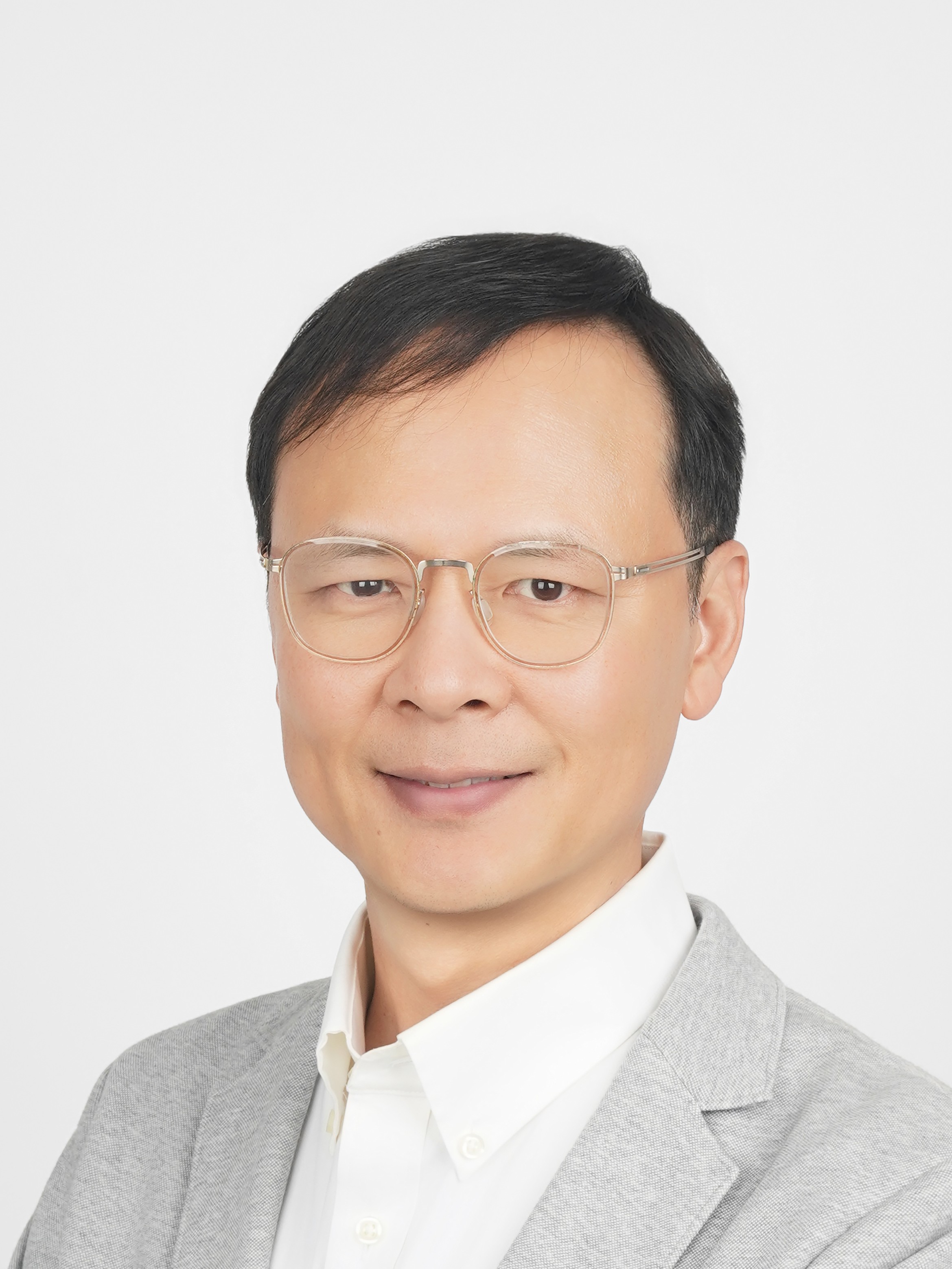}}] {Woo-Shik Kim} (Senior Member, IEEE) received the B.Eng. and M.Eng. degrees in electronics engineering from Sogang University, Seoul, South Korea, in 1999 and 2001, respectively, and the Ph.D. degree in electrical engineering from the University of Southern California, Los Angeles, CA, USA, in 2011. He began his research career at the Samsung Advanced Institute of Technology (SAIT). During his doctoral studies, he was a Research Intern with Mitsubishi Electric Research Laboratories, Dolby Laboratories, and Technicolor. Following his Ph.D., he held research positions with the Systems and Applications R\&D Center, Texas Instruments, and the Multimedia R\&D Group, Qualcomm, where he developed advanced signal processing algorithms for multimedia systems. He subsequently returned to SAIT, Samsung Electronics, where he led research projects on algorithm design for next-generation multimedia and computing devices. He is currently with Telechips, where he serves as an AI Strategist, leading research on AI applications for automotive platforms and physical AI.
Dr. Kim holds approximately 140 U.S. patents. His research interests span signal processing and computer vision, efficient and hardware-aware deep learning, and the co-optimization of algorithms and computing systems for automotive and physical AI applications.
\end{IEEEbiography}

\begin{IEEEbiography} [{\includegraphics[width=1in,height=1.25in,clip,keepaspectratio]{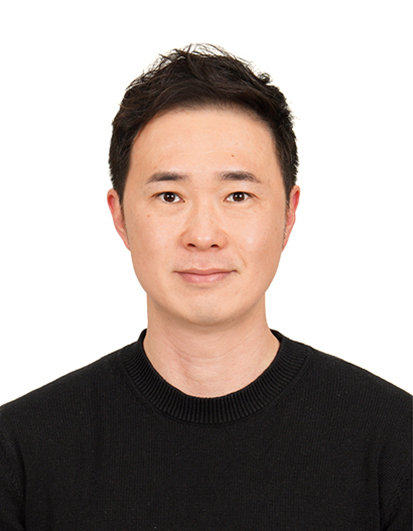}}] {Sangyoon Lee} received his Ph.D. degree in Electrical and Electronics Engineering from Yonsei University, Seoul, South Korea, in 2022. Since then, he has been with Samsung Advanced Institute of Technology. His research interests include image processing, noise analysis and modeling, noise removal, and multispectral image demosaicking.
\end{IEEEbiography}

\begin{IEEEbiography}[{\includegraphics[width=1in,height=1.25in,clip,keepaspectratio]{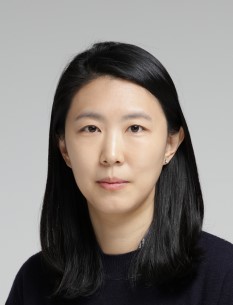}}] {YungKyung Park}, Ph.D. has been Professor at Ewha Womans University since 2012 with researching in color science field. Prior to joining Ewha Womans University, Park was senior engineer for Samsung Electronics (LCD division). During her 2 years at Samsung electronics, Park spent time doing research on Image quality and color appearance. Park received a Ph.D. in color science field from Leeds University, UK and a master degree in color imaging science from the Derby University, UK. Park received her BA and Master Degree in physics from Ewha Womans University, Korea.
\end{IEEEbiography}

\begin{IEEEbiography}[{\includegraphics[width=1in,height=1.25in,clip,keepaspectratio]{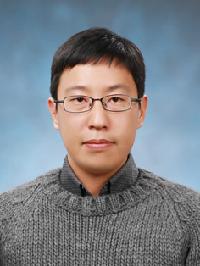}}] {Je-Won Kang} received his Ph.D degree in 2012 from the University of Southern California (USC), Los Angeles, USA. From 2011 to 2012, he was a visiting scholar at Tampere University and Nokia research center. From 2012 to 2014, he worked as a senior engineer in the Multimedia RnD and Standards team at Qualcomm, San Diego, USA, contributing significantly to international 2D and 3D video coding standards, including High Efficiency Video Coding (HEVC)/H.265, Screen Content Coding, and 3D HEVC. In 2014, he joined the Department of Electronic and Electrical Engineering at Ewha Womans University, where he now leads the Information Coding and Processing Lab as a full professor. He was honored as Ewha Fellow in 2022. He was a visiting professor in Harvard Medical School and Massachusetts General Hospital in 2021. Dr. Kang has authored over 100 peer-reviewed papers and holds more than 50 standard-essential patents. He was an APSIPA Distinguished Lecturer (2021–2022) and served as an Associate Editor for APSIPA Transactions on Signal and Information Processing. His research interests include learning-based video compression, 3D video processing, computer vision, and multimodal AI.
\end{IEEEbiography}

\end{document}